\documentclass{revtex4}
\usepackage[T1]{fontenc}
\usepackage[latin9]{inputenc}
\usepackage{amsmath}
\usepackage{amssymb}
\usepackage{graphicx}
\usepackage{esint}

\makeatletter

\providecommand{\tabularnewline}{\\}

\@ifundefined{textcolor}{}
{%
 \definecolor{BLACK}{gray}{0}
 \definecolor{WHITE}{gray}{1}
 \definecolor{RED}{rgb}{1,0,0}
 \definecolor{GREEN}{rgb}{0,1,0}
 \definecolor{BLUE}{rgb}{0,0,1}
 \definecolor{CYAN}{cmyk}{1,0,0,0}
 \definecolor{MAGENTA}{cmyk}{0,1,0,0}
 \definecolor{YELLOW}{cmyk}{0,0,1,0}
}

\makeatother

\begin{document}

\title{Constraints on a scale-dependent bias from galaxy clustering}

\author{L. Amendola$^{1}$, E. Menegoni$^{2}$, C. Di Porto$^{3,6}$, M.
Corsi$^{3}$, E. Branchini$^{3,4,5}$ }

\affiliation{$^{1}$ Institut F$\ddot{u}$r Theoretische Physik, Ruprecht-Karls-Universit$\ddot{a}$t
Heidelberg, Philosophenweg 16, 69120 Heidelberg, Germany}

\affiliation{$^{2}$ CNRS, Laboratoire Univers et Theories (LUTh), UMR 8102 CNRS,
Observatoire de Paris, Universite' Paris Diderot; 5 Place Jules Janssen,
92190 Meudon, France}

\affiliation{$^{3}$ Dipartimento di Fisica \textquotedbl{}E. Amaldi\textquotedbl{},
Universita' degli Studi \textquotedbl{}Roma Tre\textquotedbl{}, via
della Vasca Navale 84, 00146, Roma, Italy}

\affiliation{$^{4}$ INFN, Sezione di Roma Tre, via della Vasca Navale 84, I-00146
Roma, Italy}

\affiliation{$^{5}$ INAF Osservatorio Astronomico di Roma, INAF, Osservatorio
Astronomico di Roma, Monte Porzio Catone, Italy. }

\affiliation{$^{6}$ INAF Osservatorio Astronomico di Bologna, via Ranzani 1,
I-40127 Bologna, Italy}
\begin{abstract}
We forecast the future constraints on scale-dependent parametrizations
of galaxy bias and their impact on the estimate of cosmological parameters
from the power spectrum of galaxies measured in a spectroscopic redshift
survey. For the latter we assume a wide survey at relatively large
redshifts, similar to the planned Euclid survey, as baseline for future
experiments. To assess the impact of the bias we perform a Fisher
matrix analysis and we adopt two different parametrizations of scale-dependent
bias. The fiducial models for galaxy bias are calibrated using a mock
catalogs of H$\alpha$ emitting galaxies mimicking the expected properties
of the objects that will be targeted by the Euclid survey.

In our analysis we have obtained two main results. First of all, allowing
for a scale-dependent bias does not significantly increase the errors
on the other cosmological parameters apart from the \textit{rms} amplitude
of density fluctuations, $\sigma_{8}$, and the growth index $\gamma$,
whose uncertainties increase by a factor up to two, depending on the
bias model adopted. Second, we find that the accuracy in the linear
bias parameter $b_{0}$ can be estimated to within 1-2\% at various
redshifts regardless of the fiducial model. The non-linear bias parameters
have significantly large errors that depend on the model adopted.
Despite of this, in the more realistic scenarios departures from the
simple linear bias prescription can be detected with a $\sim2\,\sigma$
significance at each redshift explored. 

Finally, we use the Fisher Matrix formalism to assess the impact of assuming 
an incorrect bias model and found that the systematic errors induced on the 
cosmological parameters are similar or even larger than the statistical ones.
\end{abstract}
\maketitle

\section{Introduction}

In the next future large experiments like DESI \citep{schlegel11}
and Euclid \citep{euclidRB} will use galaxy clustering to obtain
simultaneous informations on the geometry of the Universe and the
growth rate of density fluctuations by measuring the galaxy power
spectrum or the two-point correlation function at different cosmic
epochs. These studies will be based on large surveys of extragalactic
objects and will allow to accurately estimate cosmological parameters,
among which the contribution of the Dark Energy to the cosmic density,
the nature of this elusive component and, finally, to test non-standard
theories of gravity. Since one typically observes the spatial fluctuation
in the galaxy distribution, not in the mass, some independent phenomenological
or theoretical insight of the mapping from one to the other is mandatory.
This mapping, which is commonly referred to as \textit{galaxy bias},
parametrises our ignorance on the physics of galaxy formation and
evolution and represents perhaps the most serious source of uncertainties
in the study of the large scale structure of the Universe.

Galaxy bias should not regarded as a simple {``}nuisance{''}
parameter. Its estimate is not just necessary to obtain unbiased cosmological
information. However, it also allows to discriminate among competing
models of galaxy formation and the physical processes that regulate
the evolution of stars and galaxies. Current limitations in the theoretical
models of galaxy evolution do not allow to predict galaxy bias with
an accuracy sufficient to constrain dark energy or modified gravity
models \citep{contreras013}. The alternative approach is phenomenological:
i.e. to estimate galaxy bias directly from the data.

Ref. \cite{diporto12}, among others, has shown that future galaxy
redshift surveys contain enough informations to break the degeneracy
between the galaxy bias, clustering amplitude and the growth factor,
effectively allowing to estimate galaxy bias from the data itself.
This come at the price of making some assumption on the biasing relation.
In this work we assume then that the bias relation between the galaxy
density field and the mass field is local, deterministic but not necessarily
scale-independent. The hypotheses of a local and deterministic bias
break down on galactic scales where stellar feedback processes and,
more in general, baryon physics become important. These scales are
small compared to the typical size of current and future galaxy surveys
and can be effectively smoothed out by focusing on large scale clustering
analyses.

A scale-dependent and redshift-dependent bias is a well established
observational fact. Its evidence has emerged from different type of
clustering analyses ranging from 2-point statistics \citep{norberg01, norberg02, zehavi05,coil06,bp07,nuza012,alhambra13,skibba13,marulli13,tegmark99,diporto14}
higher order statistics\citep{verde02,gaztanaga05,kayo04,nishimichi07, swanson08},
galaxy counts \citep{branchini01,marinoni05,kovac011} and gravitational
lensing \citep{hoekstra02, simon07, jullo12,comparat13}. In our analysis
we therefore assume that galaxy bias is scale and redshift dependent.
In doing so we shall adopt a parametric approach in which one assumes
\textit{a priori} a theoretically justified analytic model for the
scale dependent bias. In this approach we ignore a possible scale
dependent bias deriving from primordial non-Gaussianity \citep{dalal,grossi,theorev}
whose signature, perhaps also detectable in next generation surveys
\cite{giannantonio}, is however supposed to be seen on scales much
larger than those relevant for galaxy bias.

In this work we forecast the errors on cosmological and galaxy bias
parameters and assess the robustness of our predictions against the
choice of the bias and the fiducial model. More specifically, we consider
two different parametrizations for the scale-dependent bias: a simple
power-law model and the polynomial model proposed by \cite{cole05}.
Both provide a reasonable good fit to mock galaxies similar to those
that will be targeted by Euclid. We also adopt two different fiducial
bias models: a simple but rather unrealistic unbiased model, in which
galaxies trace mass, and a more realistic model in which the bias
parameters allow to match the 2-point clustering properties of the
mock galaxies. Finally, we restrict the analysis to large scales to
avoid strong non-linear effects, which allow us to consider linear
prescriptions complemented by some additional term to account for
mildly non-linearities.

As dataset we assume a wide spectroscopic galaxy redshift survey spanning
a large redshift range and consider, as a reference case, the upcoming
Euclid survey \cite{euclidRB}.

The layout of the paper is as follows. In section~\ref{sec:theory}
we provide the theoretical framework of our Fisher matrix analysis
and discuss the bias models used. In section~\ref{sec:Ty1} we present
the result of the analysis when galaxies are assumed to trace the
underlying mass density field at all redshifts. In section~\ref{sec:Ty2}
we adopt a more realistic fiducial model for galaxy bias, calibrated
using mock galaxy catalogs and repeat the analysis of the previous
sections. The main results are summarised and briefly discussed in
the last section.

\section{Theoretical Setup}

\label{sec:theory}

Following \cite{seo03} we model the observed galaxy power spectrum,
$P_{obs}$ at the generic redshift $z$ as: 
\begin{eqnarray}
P_{obs}(z,k) & = & G^{2}(z)b^{2}(k,z)\left(1+\frac{f(z)}{b(z,k)}\mu^{2}\right)^{2}{\frac{D_{Af}^{2}(z)H(z)}{D_{A}^{2}(z)H_{f}(z)}}P_{0f}(k)\nonumber \\
 &  & +P_{shot}(z)\label{eq:pmodel}
\end{eqnarray}
where $D_{A}$ is the angular-diameter distance, $H(z)$ is the so
called expansion history, i.e. the Hubble constant at redshift $z$,
$G(z)$ is the linear growth function normalized to unity at $z=0$,
$f(z)=d\log G/d\log a$ is the growth rate, $b(z,k)$ is the scale-dependent
bias, $P_{0}(k)$ is the matter power spectrum at the present epoch,
$\mu$ is the cosine angle between the wavenumber vector $\vec{k}$
and the line of sight direction, $P_{shot}$ is the Poisson shot noise
contribution to the power spectrum and the subscript $f$ identifies
the \textit{fiducial} model. We parametrize the growth rate as $f=\Omega_{m}^{\gamma}$
with a constant growth index $\gamma$.

We perform the forecast using the Fisher matrix information method,
i.e. by approximating the likelihood as a Gaussian in the parameters
around a particular fiducial model, i.e. a value of the parameters
that is assumed to approximate the 2-point clustering properties of
galaxies in the real Universe. Since we are mostly interested in constraining
galaxy bias, we rounded off the values of the cosmological parameters
of the fiducial model, rather than using latest experimental figures.
We set $h_{0}=0.7$, $\Omega_{m0}=0.25$; $\Omega_{b0}=0.0445$; $\Omega_{k0}=0$;
primordial slope $n_{s}=1$; and the dark energy equation of state
$w_{0}=-0.95$. Finally, we set $\gamma=0.545$ and \textit{rms} density
fluctuation at 8 $h^{-1}$ Mpc $\sigma_{8}=0.8$.

Errors in the measured spectrometric redshift, $\delta z\approx0.001(1+z)$,
propagate into errors in the estimated distances, $\sigma_{r}=\frac{\delta z}{H\left(z\right)}$.
Under the hypothesis that they are independent on the local galaxy
density we can model their effect on the estimated power spectrum
as 
\begin{equation}
P\left(z,k\right)=P_{obs}\left(z,k\right)e^{-k^{2}\mu^{2}\sigma_{r}^{2}}\,.\label{eq:ps}
\end{equation}
This term removes power on small scales and effectively damp nonlinear
effects on wavenumber larger than $\sigma_{r}^{-1}$. In order to
account for mildly nonlinear effects we follow \cite{2007ApJ...664..660E,2007ApJ...665...14S}
and multiply the power spectrum by the damping factor 
\begin{equation}
\exp\left\{ -k^{2}\left[\frac{(1-\mu^{2})\Sigma_{\perp}^{\,2}}{2}+\frac{\mu^{2}\Sigma_{\parallel}^{\,2}}{2}\right]\right\} \,.\label{eq:damping}
\end{equation}
This elliptical Gaussian function models the displacement field in
Lagrangian space on scales $\gtrsim10\,h^{-1}$Mpc, where we focus
our analysis. In this expression $\Sigma_{\perp}$ and $\Sigma_{\parallel}$
represent the displacement across and along the line of sight, respectively.
They are related to the growth factor $G$ and to the growth rate
$f$ through $\Sigma_{\perp}=\Sigma_{0}G$ and $\Sigma_{\parallel}=\Sigma_{0}G(1+f)$.
The value of $\Sigma_{0}$ is proportional to $\sigma_{8}$. For our
fiducial cosmology where $\sigma_{8}=0.8$~, we have $\Sigma_{0}=11\,h^{-1}$
Mpc. This implies that power is removed below this scale.

Assuming that the fluctuation Fourier modes are Gaussian variates,
the Fisher matrix at each redshift shell is \citep{ewt98,teg97} 
\begin{equation}
F_{ij}=2\pi\int_{k_{min}}^{k_{max}}\frac{\partial\log P\left(k_{n}\right)}{\partial\theta_{i}}\frac{\partial\log P\left(k_{n}\right)}{\partial\theta_{j}}\cdot V_{eff}\cdot\frac{k^{2}}{8\pi^{3}}\cdot dk\label{eq:FisherMatrix}
\end{equation}
where the derivatives are evaluated at the parameter values of the
fiducial model. Here, the maximum frequency $k_{max}(z)$ is set by
the scale at which fluctuations grow nonlinearly while $k_{min}(z)$
by the largest scale that can be observed in the given redshift shell.
We set a hard small-scale cut-off $k_{max}=0.5\,h^{-1}$ Mpc at all
redshifts which, together with the damping terms~(\ref{eq:ps}) and
(\ref{eq:damping}) account for non linearities. On large scale we
set $k_{min}=0.001\,h^{-1}$ Mpc. However, its precise value is not
very relevant since the contribution of low $k$ modes to the Fisher
matrix is negligible. $V_{\mathrm{eff}}$ indicates the effective
volume of the survey defined as: 
\begin{eqnarray}
V_{\mathrm{eff}} & \equiv & \int\left[\frac{n\left(\vec{r}\right)P\left(k,\mu\right)}{n\left(\vec{r}\right)P\left(k,\mu\right)+1}\right]^{2}d\vec{r}=\left[\frac{nP\left(k,\mu\right)}{nP\left(k,\mu\right)+1}\right]^{2}V_{survey}\,\label{eq:Volume}
\end{eqnarray}
where $n=n(z)$ is the galaxy density at redshift $z$. The second
equality in Equation~(\ref{eq:Volume}) holds if the co-moving number
density is constant within the volume considered. This assumption,
which we adopt in our analysis, is approximately true in a sufficiently
narrow range of redshifts. For this reason, we perform the Fisher
matrix analysis in different, non-overlapping redshift bins listed
in Table~\ref{tab:z-density}, together with their mean galaxy number
density. The redshift range, the size of the bin and the number density
of objects roughly match the analogous quantities that are expected
in the Euclid spectroscopic survey. To improve the correspondency,
we multiply the galaxy number densities in Table~\ref{tab:z-density}
by an \textquotedbl{}efficiency\textquotedbl{} factor 0.5 and assumes
a survey area of 15,000 deg$^{2}$.

For additional robustness of our results, we marginalize over the
Alcock-Paczinsky parameters. That is, we convert the wavenumber norm
$k$ and direction cosine $\mu$ from the fiducial to the any other
cosmology using a free Hubble function and angular diameter distance
parameters and marginalize over them, instead of projecting over the
background parameters $\Omega_{m0},h_{0}$.

\begin{table}[htbp]
\centering \protect\protect\protect\protect\caption{ Redshift bins used in our analysis and their mean galaxy number density.
Col. 1: central redshift of each redshift shell with width $\Delta z=0.2$.
Col 2: Mean number density of objects in $h^{3}{\rm Mpc}^{-3}$. These
numbers match those expected for a Euclid-like survey according to
\cite{euclidRB}. }

\begin{tabular}{lccc}
\hline 
\multicolumn{1}{c}{$z$} & $n_{dens}$  &  & \tabularnewline
\hline 
$0.6$  & $3.56\times10^{-3}$  &  & \tabularnewline
$0.8$  & $2.42\times10^{-3}$  &  & \tabularnewline
$1.0$  & $1.81\times10^{-3}$  &  & \tabularnewline
$1.2$  & $1.44\times10^{-3}$  &  & \tabularnewline
$1.4$  & $0.99\times10^{-3}$  &  & \tabularnewline
$1.6$  & $0.55\times10^{-3}$  &  & \tabularnewline
$1.8$  & $0.29\times10^{-3}$  &  & \tabularnewline
$2.0$  & $0.15\times10^{-3}$  &  & \tabularnewline
\hline 
\end{tabular}\label{tab:z-density} 
\end{table}

\subsection{Analytic models for scale-dependent bias}

\label{sec:biases}

The final ingredient in Eq.~(\ref{eq:pmodel}) and the focus of this
paper is the scale-dependent galaxy bias $b(z,k)$. Several authors
have proposed different models, both phenomenological and theoretical
\cite{seljak01,cole05,seo05,sw06,huff07,smith07}. In this work we
are not too concerned on the accuracy of bias models. Our goal is
to assess the impact of a scale-dependent galaxy bias in the analysis
of future galaxy surveys. For this purpose we have decided to adopt
two rather simple models, the Power Law and the Q-Model, that nonetheless
provide a good match to the galaxy bias measured in numerical experiments,
as we shall see. The reason for choosing these models is twofold.
First, they have been already used in the literature, making it possible
to compare our results with existing ones and use previous results
to set the range in which the model parameters can vary. Second, their
simple form allows us to compute the power spectrum derivatives in
the Fisher matrix analytically.

The Power Law bias model has the form \cite{fg93}: 
\begin{equation}
b(z,k)=b_{0}(z)+b_{1}(z)\left(\frac{k}{k_{1}}\right)^{n}\,,\label{eq:plmodel}
\end{equation}
where the pivot scale $k_{1}$ is introduced only to deal with dimensionless
parameters. Its value does not impact on our analysis and, without
lack of generality, we set $k_{1}=1$ $h$ Mpc$^{-1}$. The slope
$n$ is not treated as a free parameter but is kept fixed. However,
to check the sensitivity of our results on the power law index we
have considered three different values: $n=1\,,1.28,\,2$. As we shall
see the value $n=1.28$ corresponds to the one that provides the best
fit to the bias of mock galaxies measured in simulated catalogs. $n=1$
also provides an acceptable fit to the mock galaxy bias. The case
$n=2$ should be regarded as an extreme case since it provides a poor
fit to the simulated data both at large and at small scales. We note
that the Power Law model is similar to the one proposed by \cite{seo05}
in those $k-$ranges in which the power spectrum can be approximated
by a power law.

The Q-Model is also phenomenological. It has been proposed by \cite{cole05}
from the analysis of mock halo and galaxy catalogs extracted from
the Hubble volume simulation. Its form is 
\begin{equation}
b(z,k)=b_{0}(z)\left[\frac{1+Q(z)(k/k_{1})^{2}}{1+A(z)(k/k_{1})}\right]^{1/2}\,,\label{eq:qmodel}
\end{equation}
In our analysis all three parameters $b_{0}$, $Q$ and $A$ are free
to vary in each redshift bin. Therefore, the Q-Model has additional
degrees of freedom with respect to the Power Law model.

Finally, we need to specify the parameters of the fiducial model.
In this work we consider two different fiducial models, Type 1 and
Type 2, corresponding to two different choices of parameters for each
bias model, totalling to four fiducial models. Type 1 models (denoted
as FM1-PL for the power law bias and FM1-Q for the Q-model) represent
the simple but rather unphysical case of galaxy tracing mass at all
redshifts. Let us stress that assuming a scale-independent \emph{fiducial}
model does not imply that we are assuming a scale-independent bias
since we differentiate the power spectrum in the Fisher matrix with
respect to the scale-dependent bias coefficients: $b_{1},\,Q,\,A$.
Instead, assuming a scale-independent fiducial model only implies
that in Type 1 models the derivative is evaluated at the fiducial
values $b_{1},\,Q\,{\rm and}\,A=0$. The parameters that identify
the fiducial models are listed in Table~\ref{tab:T1}. Note that
we also consider for comparison the case of scale-independent bias
(first row). This case is identical to choosing $n=0$ in the power
law model, i.e. to $b(z)=b_{0}(z)+b_{1}(z)$, so we will refer to
this case as $n=0$ fiducial.

Type 2 models (denoted as FM2-PL and FM2-Q) are more realistic. The
fiducial parameters of the Power Law and Q-Models were determined
by matching the bias of mock H$\alpha$ emitting galaxies in the simulations
described in Section~\ref{sec:mocks}. Table~\ref{tab:T2} lists
the parameters of the fiducial models.

Note that for the power law cases the values of the parameters depend
on the choice of the power law index $n$ so that, since we explore
the cases $n=1\,,1.28,\,2$, we end up by having several Power Law
fiducial models.

\begin{table}
\protect\protect\protect\caption{Bias parameters for Type 1 fiducial models.}

\begin{centering}
\begin{tabular}{c|c|cccc|ccc}
\hline 
 & \multicolumn{1}{c|}{} & \multicolumn{4}{c|}{FM1-PL} & \multicolumn{3}{c}{FM1-Q}\tabularnewline
 & \multicolumn{1}{c|}{$n=0$} & \multicolumn{2}{c}{$n=1$} & \multicolumn{2}{c}{$n=2$ } &  &  & \tabularnewline
\hline 
\hline 
$z$  & $b_{0}$  & $b_{0}$  & $b_{1}$  & $b_{0}$  & $b_{1}$  & $b_{0}$  & $Q$  & $A$ \tabularnewline
\hline 
all  & $1$  & $1$  & $0$  & $1$  & $0$  & $1$  & $0$  & $0$ \tabularnewline
\hline 
\end{tabular}
\par\end{centering}
\label{tab:T1} 
\end{table}

\begin{table}[htbp]
\centering \protect\protect\protect\protect\caption{Bias parameters for Type 2 fiducial models.}

\centering{}%
\begin{tabular}{c|cccccc|ccc}
\hline 
 & \multicolumn{6}{c|}{FM2-PL} & \multicolumn{3}{c}{FM2-Q}\tabularnewline
 & \multicolumn{2}{c}{$n=1$} & \multicolumn{2}{c}{$n=1.28$ } & \multicolumn{2}{c}{$n=2$ } &  &  & \tabularnewline
$z$  & $b_{0}$  & $b_{1}$  & $b_{0}$  & $b_{1}$  & $b_{0}$  & $b_{1}$  & $b_{0}$  & $A$  & $Q$ \tabularnewline
\hline 
\hline 
$0.8$  & 1.04  & 0.67  & 1.09  & 0.66  & 1.17  & 0.68  & 1.26  & 1.7  & 4.54 \tabularnewline
$1.0$  & 1.13  & 0.74  & 1.19  & 0.75  & 1.28  & 0.79  & 1.36  & 1.7  & 4.92 \tabularnewline
$1.2$  & 1.22  & 0.99  & 1.30  & 0.97  & 1.41  & 1.02  & 1.49  & 1.7  & 5.50 \tabularnewline
$1.4$  & 1.36  & 1.09  & 1.44  & 1.06  & 1.55  & 1.12  & 1.63  & 1.7  & 5.70 \tabularnewline
$1.6$  & 1.49  & 1.22  & 1.58  & 1.19  & 1.71  & 1.25  & 1.75  & 1.7  & 6.62\tabularnewline
$1.8$  & 1.61  & 1.40  & 1.72  & 1.40  & 1.88  & 1.44  & 1.92  & 1.7  & 6.99 \tabularnewline
\hline 
\end{tabular}\label{tab:T2} 
\end{table}

The complete set of parameters that are free to vary is then $h,\Omega_{b0},\Omega_{m0},\Omega_{k0},n_{s},\gamma,\sigma_{8}$
plus, for each redshift shell, the shot noise $P_{noise}$ and $b_{0},b_{1}$
or alternatively $b_{0},A,Q$. This amounts to a total of 39 independent
parameters when we adopt 8 redshift bins. All the other parameters
(the bias power law index $n$, the damping model, the survey parameters
like galaxy density, volume, redshift error as well as the dark energy
parameters $w_{0}$ and $w_{1}$ are kept fixed).

\subsection{Derivatives in the Fisher Matrix analysis}

In order to evaluate the Fisher matrix we compute the derivatives
of the power spectrum in Eq.~(\ref{eq:pmodel}) with respect to the
parameters on the fiducial model. The $1\sigma$ error for each parameter
of the model, $p_{i}$, is $\sigma_{p_{i}}=\sqrt{(F^{-1})_{ii}}$,
where $F^{-1}$ is the inverse Fisher matrix.

Focusing on the free parameters of the biasing function we have for
the Power Law bias model: 
\begin{eqnarray}
\frac{d\ln P}{db_{0}}|_{f} & = & \frac{2}{b_{0}^{f}}-\frac{2f\mu^{2}}{b_{0}^{f}(b_{0}^{f}+f\mu^{2})}\\
\frac{d\ln P}{db_{1}}|_{f} & = & \frac{2}{b_{0}^{f}}(k/k_{1})^{n}-\frac{2f\mu^{2}(k/k_{1})^{n}}{b_{0}^{f}(b_{0}^{f}+f\mu^{2})}
\end{eqnarray}
For the Q-model we have instead : 
\begin{equation}
\frac{d\ln P}{db_{0}}|_{f}=\frac{2}{b_{0}^{f}}-\frac{2f\mu^{2}}{b_{0}^{f}\left[b_{0}^{f}\left(\frac{1+Q_{f}(k/k_{1})^{2}}{1+A_{f}(k/k_{1})}\right)^{1/2}+f\mu^{2}\right]}
\end{equation}
and 
\begin{eqnarray}
\frac{d\ln P}{dQ}|_{f} & = & \frac{(k/k_{1})^{2}}{1+Q_{f}(k/k_{1})^{2}}-\frac{f\mu^{2}(k/k_{1})^{2}}{1+Q_{f}(k/k_{1})^{2}}\frac{1}{f\mu^{2}+b_{0}^{f}\left[\frac{1+Q_{f}(k/k_{1})^{2}}{1+A_{f}(k/k_{1})}\right]^{1/2}}\\
\frac{d\ln P}{dA}|_{f} & = & -\frac{k/k_{1}}{1+A_{f}(k/k_{1})}+\frac{f\mu^{2}(k/k_{1})b_{0}^{f}}{[1+A_{f}(k/k_{1})]\left[f\mu^{2}+b_{0}^{f}\left[\frac{1+Q_{f}(k/k_{1})^{2}}{1+A_{f}(k/k_{1})}\right]^{1/2}\right]}
\end{eqnarray}

\section{\label{sec:Ty1}Type 1 Fiducial Models. Results}

In this Section we show the results of the Fisher Matrix analysis
performed using Type 1 models FM1-PL and FM1-Q. They represent the
case of a survey of objects that trace the underlying mass density
field in an unbiased way.

\subsection{Power law case}

For the power law model we have explored two cases corresponding to
different choices of the power law index: $n=1$ and $n=2$. The more
realistic case $n=1.28$ will be considered in the next Section. In
addition, we consider the case $n=0$ that represents the scale-independent
hypothesis often assumed in many Fisher matrix forecast papers. For
this particular case we do not compute derivatives with respect to
$b_{1}$. The parameters of the fiducial models are reported in Table~\ref{tab:T1}.

The goal of our analysis is twofold: to assess the impact of a scale-dependent
bias on the measurement of the cosmological parameters and to estimate
the accuracy with which we can measure the parameters that characterise
the bias. Let us first focus on the first task.

\begin{figure}
\includegraphics[width=0.5\textwidth]{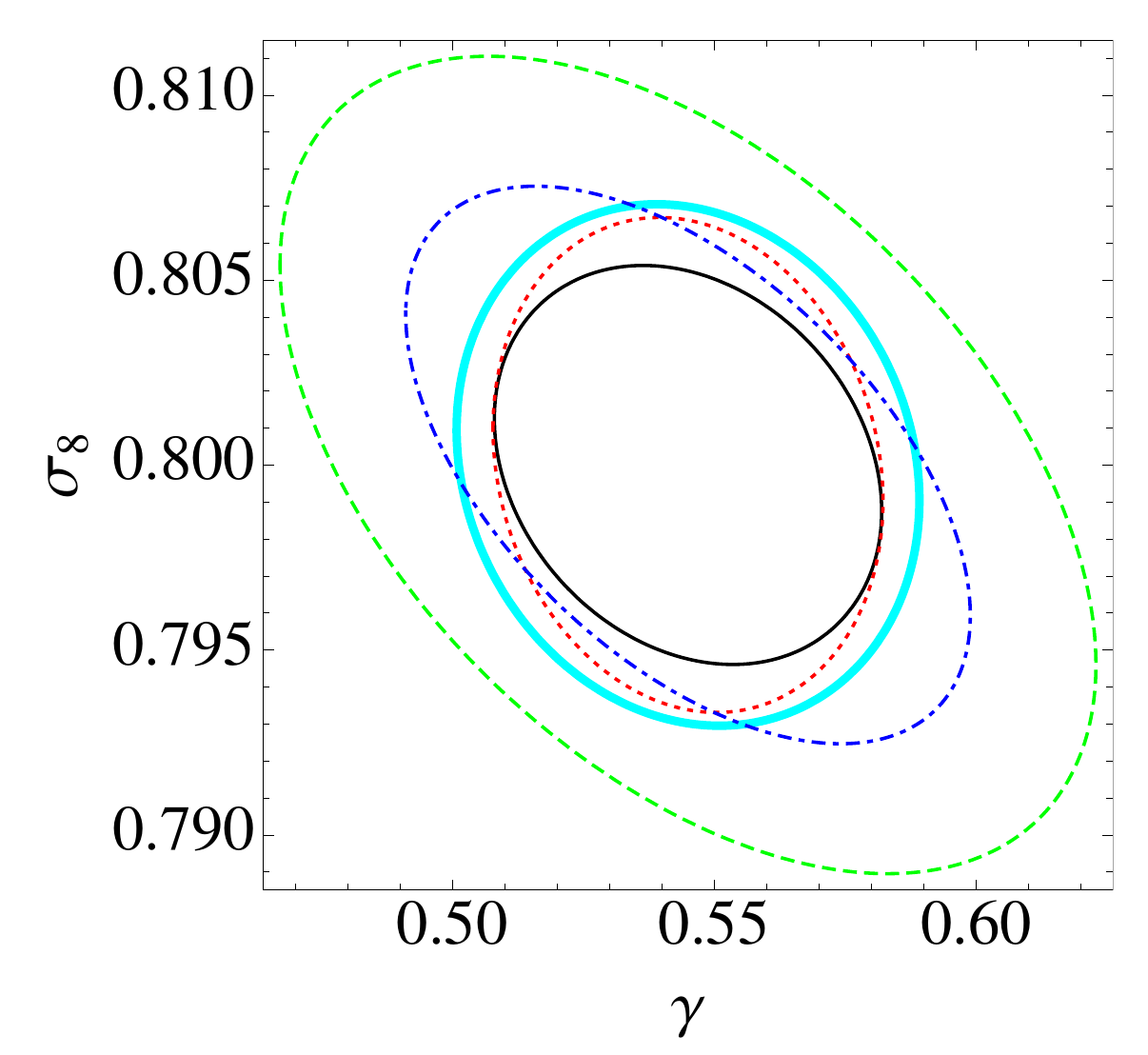}
\protect\protect\protect\protect\caption{\label{fig:s8gamma} 68 \% probability contours for $\sigma_{8}$
and $\gamma$. Black, continuous: standard scale-independent case,
i.e. $n=0$. Red, Dotted: FM1-PL with $n=1$. Cyan, continuous: FM1-Q.
Blue, Dot-Dashed: FM2-PL with $n=1$. Green, dashed: FM2-Q. }
\end{figure}

The $1\sigma$ errors on the cosmological parameters are listed in
Table \ref{tab:a-referencecosmoerrors}. Unless otherwise specified,
the quoted errors are always obtained after marginalising over all
the other parameters. For all parameters except the mass variance
$\sigma_{8}$ and the growth index $\gamma$ the errors are largely
independent from $n$. In fact in most cases they slightly decrease
when the scale dependence is stronger. The values of $\sigma_{8}$
and $\gamma$ show the opposite trend, although the effect is quite
small (below 10 \%). We conclude that allowing for a scale dependent
bias has little effect on the precision in which we can measure most
cosmological parameters. It is interesting to note that the accuracy
of the growth rate $\gamma$ is 4-5\% when marginalising over all
parameters, including the scale and redshift-dependent bias. Figure~\ref{fig:s8gamma}
gives a visual impression of this fact. It shows the 1-$\sigma$ likelihood
contours for $\sigma_{8}$ and $\gamma$ for most of the fiducial
models explored in this paper. The likelihood ellipse obtained in
the FM1-PL case when $n=1$ (red, dotted curve) is only slightly larger
than that corresponding to a scale-independent bias (continuous, black
line). The likelihood contours obtained for $n=2$, not plotted to
avoid overcrowding, are also similar. We further notice that there
is little correlation between $\sigma_{8}$ and $\gamma$.

\begin{table}[htbp]
\centering \protect\protect\protect\protect\caption{\label{tab:a-referencecosmoerrors} $1\sigma$ errors on cosmological
parameters for Type 1 fiducial models.}

\begin{tabular}{l|c|cc|c}
\hline 
 &  & \multicolumn{2}{c|}{FM1-PL} & \multicolumn{1}{c}{FM1-Q}\tabularnewline
Error  & $n=0$  & $n=1$  & $n=2$  & \tabularnewline
\hline 
$\sigma_{h}$  & 0.036  & 0.038  & 0.037  & 0.039 \tabularnewline
$\sigma_{\Omega_{m}h^{2}}$  & 0.015  & 0.016  & 0.015  & 0.016 \tabularnewline
$\sigma_{\Omega_{b}h^{2}}$  & 0.0034  & 0.0036  & 0.0034  & 0.0036 \tabularnewline
$\sigma_{n_{s}}$  & 0.036  & 0.042  & 0.036  & 0.044 \tabularnewline
$\sigma_{\gamma}$  & 0.024  & 0.025  & 0.028  & 0.029 \tabularnewline
$\sigma_{\sigma_{8}}$  & 0.0036  & 0.0044  & 0.0045  & 0.0047 \tabularnewline
\hline 
\end{tabular}
\end{table}

%\begin{table}[htbp]
%\centering \protect\protect\protect\protect\caption{ Errors bias parameters $b_{0}$ with $n=0$. }

%\begin{tabular}{lccc}
%\hline 
%\multicolumn{1}{c}{$z$} & $\sigma_{b_{0}}$  &  & \tabularnewline
%\hline 
%$0.6$  & $0.007$  &  & \tabularnewline
%$0.8$  & $0.008$  &  & \tabularnewline
%$1.0$  & $0.009$  &  & \tabularnewline
%$1.2$  & $ 0.0101$  &  & \tabularnewline
%$1.4$  & $0.011$  &  & \tabularnewline
%$1.6$  & $0.012$  &  & \tabularnewline
%$1.8$  & $0.014$  &  & \tabularnewline
%$2.0$  & $0.018$  &  & \tabularnewline
%\hline 
%\end{tabular}\label{tab:errors-b0-n0} 
%\end{table}

\begin{table}[htbp]
\centering \protect\protect\protect\protect\caption{\label{tab:a-biaserrorsb0b1nref1} Errors on bias parameters for Type
1 fiducial models.}

\begin{tabular}{c|c|cccc|ccc}
\hline 
 & \multicolumn{1}{c|}{} & \multicolumn{4}{c|}{FM1-PL} & \multicolumn{3}{c}{FM1-Q}\tabularnewline
 & \multicolumn{1}{c|}{$n=0$} & \multicolumn{2}{c}{$n=1$} & \multicolumn{2}{c|}{$n=2$} & \multicolumn{3}{c}{}\tabularnewline
$z$  & $\sigma_{b_{0}}$  & $\sigma_{b_{0}}$  & $\sigma_{b_{1}}$  & $\sigma_{b_{0}}$  & $\sigma_{b_{1}}$  & $\sigma_{b_{0}}$  & $\sigma_{Q}$  & $\sigma_{A}$ \tabularnewline
\hline 
\hline 
$0.6$  & 0.007  & 0.013  & 0.14  & 0.0081  & 1.2  & 0.017  & 3.04  & 0.35 \tabularnewline
$0.8$  & 0.008  & 0.013  & 0.13  & 0.0093  & 0.97  & 0.017  & 2.5  & 0.32 \tabularnewline
$1.0$  & 0.009  & 0.013  & 0.12  & 0.011  & 0.86  & 0.017  & 2.2  & 0.31 \tabularnewline
$1.2$  & 0.010  & 0.014  & 0.12  & 0.012  & 0.82  & 0.018  & 2.2  & 0.31 \tabularnewline
$1.4$  & 0.011  & 0.014  & 0.13  & 0.013  & 0.91  & 0.019  & 2.5  & 0.34 \tabularnewline
$1.6$  & 0.012  & 0.016  & 0.16  & 0.014  & 1.2  & 0.023  & 3.4  & 0.42 \tabularnewline
$1.8$  & 0.014  & 0.019  & 0.22  & 0.016  & 1.9  & 0.027  & 5.4  & 0.59 \tabularnewline
$2.0$  & 0.018  & 0.026  & 0.34  & 0.019  & 3.3  & 0.037  & 9.4  & 0.97 \tabularnewline
\hline 
\end{tabular}
\end{table}

The expected $1\sigma$ errors for the bias parameters $b_{0}$ and
$b_{1}$ are listed in Table~\ref{tab:a-biaserrorsb0b1nref1} for
the cases $n=0,\,1\,$ and $2$. Errors on $b_{1}$ are larger than
those on $b_{0}$ and their size increase with $n$. When $n=2$ they
are ten times larger than with $n=1$. On the contrary, errors on
$b_{0}$ weakly depend on $n$. This is not surprising since $b_{1}$
is constrained by the power spectrum behaviour at high $k$, the larger
the value of $n$ the larger the values of $k$, where our analysis
is less sensitive due to the damping terms and the hard $k_{max}$
cut. A second trend is with the redshifts: errors on the bias parameters
increase with the redshift, irrespective of the $n$ value. Again,
this is not surprising since it merely reflects the fact that the
effective volume of the survey monotonically decreases when moving
to high redshifts due to the smaller galaxy densities.

When compared to the results of Ref. \cite{diPorto}, in which bias
was assumed to be scale-independent, we notice that our constraints
on $b_{0}$ are twice weaker than their \textquotedbl{}optimistic,
internal bias\textquotedbl{} case at $z=1.8$ and $z=2.0$. This quantifies
the effect of allowing for an additional degree of freedom, the scale
dependent bias, represented by the new parameter $b_{1}$.

\begin{figure}[htb!]
\centering{}\hspace*{-1cm} %
\begin{tabular}{cc}
\includegraphics[width=0.45\textwidth]{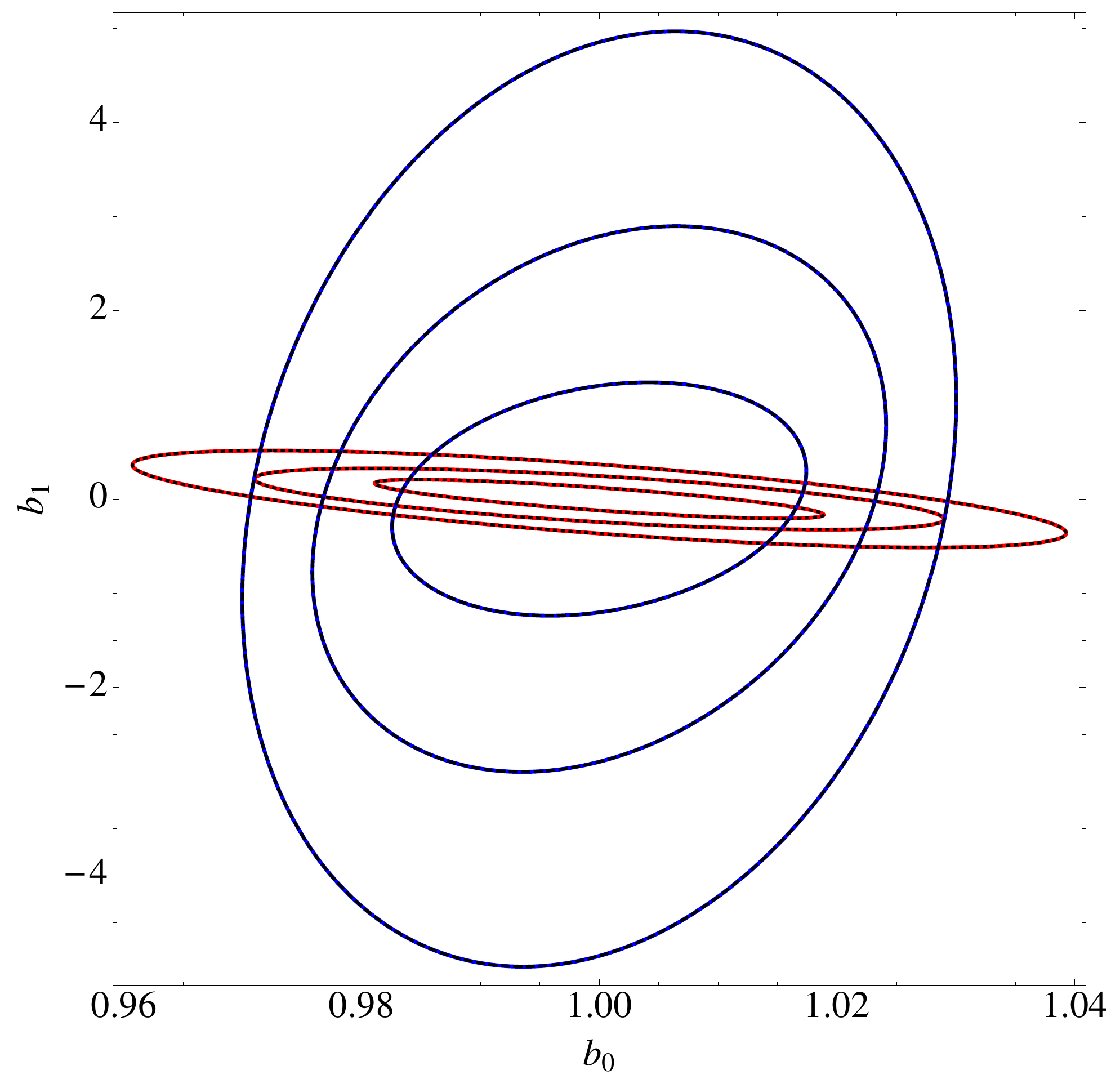}  & \tabularnewline
\end{tabular}\protect\caption{\label{fig:a-fidnref1} $68\%$ probability contours for the parameters
$b_{0}$ and $b_{1}$ of the FM1-PL model. The dotted red line is
the case with $n=1$ while the dashed line in blue shows the case
$n=2.$ The redshift bins are $z=0.6,1.8,2.0$, from inside out.}
\end{figure}

In Figure~\ref{fig:a-fidnref1} we show the 68 \% probability contours
in the $b_{0}$-$b_{1}$ plane for $n=1$ and $n=2$, respectively.
Larger ellipses refer to higher redshift bins. For the case $n=1$
there is a strong anti-correlation between $b_{0}$ and $b_{1}$ which
stems from the fact that an increase in the linear bias term $b_{0}$
can be partially compensated by reducing the amplitude of the scale-dependent
term $b_{1}$. Increasing the scale dependency, i.e. setting $n=2$
reduces the correlation between $b_{0}$ and $b_{1}$. This is due
to the fact that a strong scale dependent bias has little impact on
large $(k\ll k_{1})$ scales and therefore cannot effectively compensate
a variation of the linear bias on the scales that are relevant for
our analysis.

\subsection{Q-model}

This version of Type 1 fiducial model is characterised by three parameters,
$b_{0}$, $A$ and $Q$, rather than two. We have repeated the same
analysis performed as in the FM1-PL case and summarised the results
in Tables~ \ref{tab:a-referencecosmoerrors} and \ref{tab:a-biaserrorsb0b1nref1}.

The errors on the cosmological parameters are very similar to those
obtained in the FM1-PL case, confirming that the accuracy in the estimate
of the cosmological parameters is little affected by the adoption
of a scale-dependent bias model, even when we introduce an additional
degree of freedom. For this reason the 68\% error contours for $\sigma_{8}$
and $\gamma$ (cyan thick curve in Figure~\ref{fig:s8gamma}) are
quite close to that of the FM1-PL $n=1$ case (red, dotted curve).

\begin{figure}[htb!]
\centering{}\hspace*{-1cm} %
\begin{tabular}{cc}
\includegraphics[width=0.45\textwidth]{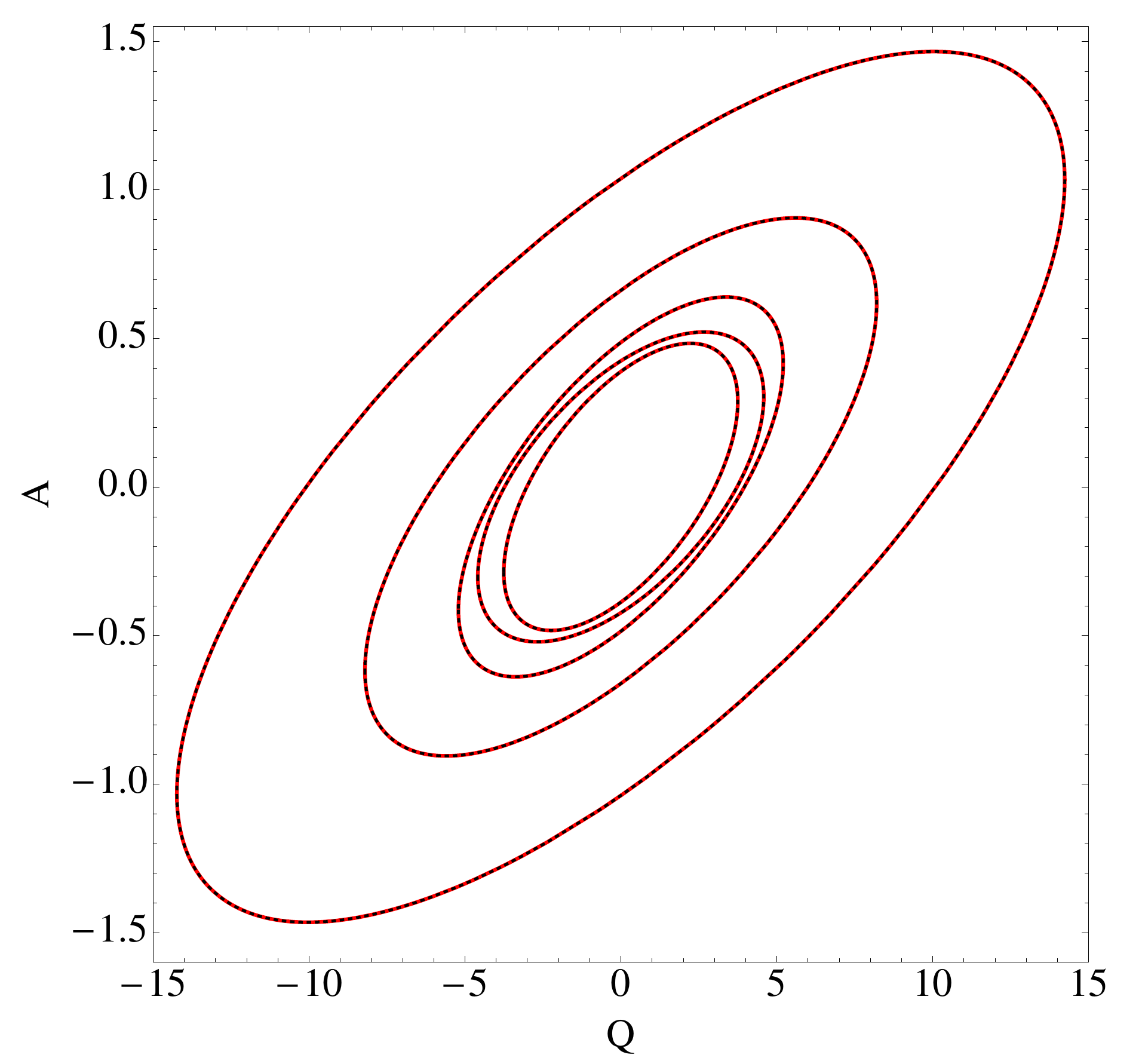}  & \tabularnewline
\end{tabular}\protect\caption{ Contours plots for 68 \% probability contours for the parameters
$A$ and $Q$ of the FM1-Q model. We decided to plot only the bins
$z=0.6,0.8,1.6,1.8,2.0$, from inside out, to improve clarity.\label{fig:a-fidAQ}}
\end{figure}

Errors on the linear bias parameters $b_{0}$ (Table~\ref{tab:a-biaserrorsb0b1nref1})
are small, with a magnitude similar to that of the FM1-PL case with
$n=1$. On the contrary, the errors on $A$ and $Q$ are quite large,
although we cannot directly compare their size to the errors on $b_{1}$.
This is not entirely surprising: it is the effect of having one more
parameter to marginalize over. To further investigate the possible
degeneracy among the bias parameters we plot the 68 \% uncertainty
contours in the $A$-$Q$ plane in Fig.~\ref{fig:a-fidAQ}. The size
of the errors, and consequently the area of the corresponding ellipse
increases with the redshift. They are positively correlated and the
strength of the correlation also increases with the redshift.

\section{\label{sec:Ty2}Type 2 fiducial models. Results}

In this Section we repeat the Fisher matrix analysis performed in
the previous Section using the same models for the power spectrum
but assuming a Type 2 fiducial model for the bias. The parameters
of the fiducial models have been provided by the best fit to the bias
of the simulated H$\alpha$ galaxies that will be targeted by the
Euclid spectroscopic redshift survey. According to Ref. \cite{euclidRB}
Euclid will produce a redshift catalog of H$\alpha$-line emitting
galaxies with line flux above H$\alpha_{{\rm min}}=3\cdot10^{-16}$
erg cm$^{-2}$ s$^{-1}$ with high completeness and purity. Ideally,
one would fix the parameters of the fiducial model by matching observations.
Unfortunately, the samples currently available are too small to reliably
estimate the bias of high redshift, H$\alpha$ galaxies. For this
reason we consider instead a catalog of simulated H$\alpha$ galaxies
designed to mimic the predicted properties of the objects that will
be observed by Euclid and measure their bias from the clustering properties.
The detailed procedure is described in the next Section.

Another important difference with Section~\ref{sec:Ty2} is that
we now consider a smaller redshift range ($z=[0.8,1.8]$). This is
a conservative choice mainly driven by the goal of matching as close
as possible the range that will be best exploited for scientific analyses
in the Euclid survey. %and partially
%determined by the characteristics of the mock catalogs available.

\subsection{Galaxy bias from mock catalogs}

\label{sec:mocks}

We consider the so-called \textquotedbl{}100 deg$^{2}$ light-cone\textquotedbl{}
mock galaxy catalog \footnote{The catalogs are publicly available at http://astro.dur.ac.uk/$\sim$d40qra/lightcones/EUCLID/ }
obtained by applying the semi-analytic galaxy formation model \texttt{GALFORM}
\citep{bower06} to the outputs of the Millennium $N$-body simulation
\citep{springel05}. The simulated volume consists of a light-cone
with a sky coverage of 100.206 deg$^{2}$ spanning the redshift range
$[0.0,2.0]$. Mock galaxies are characterised by angular position,
redshift, co-moving distances and a number of observational properties
among which the luminosity of the H$\alpha$ line. From these mocks
we selected a subsample of objects in the range $z=[0.6,2.0]$ and
with H$\alpha$ line flux larger than H$\alpha_{{\rm min}}$.

To estimate galaxy bias we have compared the power spectrum of the
H$\alpha$ galaxies to that of the linear mass power spectrum obtained
from \texttt{CAMB} \citep{lewis00} in six redshift bin of width $\Delta z=0.2$.
The galaxy bias and its dependence on $k$ is obtained from the ratio
of these two quantities. This bias estimate is quite noisy due to
the limited number density of mock galaxies. To regularise its behaviour
we fit the Power Law and Q-model and determine the parameters of the
fiducial by minimising the $\chi^{2}$ difference between model and
estimated bias. Dealing with analytic expression for the bias also
facilitates its implementation in the Fisher matrix analysis.

The detailed procedure to determine the realistic model the galaxy
bias is as follows: 
\begin{itemize}
\item We extract six, partially overlapping cubic boxes from the simulated
light cone. These boxes are fully contained in the light cone, aligned
along the line of sight and centred at the redshifts indicated in
Table ~\ref{tab:T2}. Their size increases with the redshift. 
\item We select galaxies brighter than H$\alpha_{{\rm min}}$ in each box
and compute a statistical weight to account for the selection implied
by the flux cut. 
\item We use a Fast Fourier Transform based estimator, similar to \citep{fkp},
to measure the power spectrum of the mock galaxies within each cube.
We ignore the effect of peculiar velocities and limit our analysis
between $k_{{\rm max}}=1h\,{\rm Mpc}^{-1}$ and $k_{{\rm min}}$.
The value of $k_{{\rm max}}$ is chosen to minimise the impact of
shot noise and aliasing. The value of $k_{{\rm min}}$ depends on
the redshift and is determined by the size of each box. 
\item We estimate the scale dependent galaxy bias from the ratio between
the mock galaxy power spectrum %$P_{{\rm Lin,M}}(k,z)$ 
and the \texttt{CAMB} power spectrum of the mass obtained using the
same cosmological parameters as the Millennium simulation. 
\item We fit the power law model and the Q-model to the estimated mock galaxy
bias, $b(k,z)$, by minimising the $\chi^{2}$ of the residuals. In
the procedure we assume that the errors on the estimated power spectrum
are the same as of the power spectrum estimator computed in \citep{fkp}.
In the $\chi^{2}$ minimisation procedure we fixed some of the bias
parameters. For the Power Law model we set $n=1,\,1.28$ and $2$.
%($n=1.28$ being the results of the
%best fit with all three parameters free) 
For the Q-model we set $A=1.7$ according to Ref. \cite{cole05}.
The minimisation is first carried out over a conservative range of
wave numbers, $[k_{{\rm min}},k_{{\rm max}}]=[0.03,0.3]$, that is
gradually increased in both direction and stop when $\chi^{2}/{\rm d.o.f.}\sim1$.
Note that the value $n=1.28$ is the one that provides the best fit
to the measured bias when $n$ is also free to vary. 
\end{itemize}
The parameters of the fiducial models obtained from the $\chi^{2}$
fit are listed in Tab.~\ref{tab:T2}. We reiterate that for the Power
Law model the best fit to the data is obtained with $n=1.28$. The
quality of the fit is still acceptable for $n=1$ whereas $n=2$ provides
a poor fit both at small and large $k$-values. For this reason the
case $n=2$ should be regarded as an extreme and somewhat unrealistic
case.

\subsection{Power Law case}

Table~\ref{tab:t2cosmoerr} shows the $1\sigma$ errors on the cosmological
parameters when one assumes FM2-PL as a fiducial model for galaxy
bias. Errors are similar to those of the Type 1 model for all parameters
except for $\sigma_{8}$ and $\gamma$ They show little or no dependence
on the power law index $n$. Errors on $\sigma_{8}$ and $\gamma$
are significantly larger since these parameters are partially degenerate
with the bias. Indeed, the amplitude of the power spectrum is sensitive
to the combination $\sigma_{8}b(k,z)$ and the strength of the redshift
distortion to $f/b(k,z)$. In addition $\sigma_{8}$ and $\gamma$
are significantly anti-correlated (solid ellipse in Fig.~\ref{fig:s8gamma}).
This anti-correlation reflects the fact that an increase in amplitude
of the mass power spectrum (i.e. an increase in $\sigma_{8}$) can
be compensated by a decrease in the growth rate $\Omega_{m}^{\gamma}$.
The reason why this anti-correlation was not seen in the Type 1 model
case is that in this new fiducial model the bias is significantly
larger than unity. As a consequence a much larger increase in the
growth rate is now required to compensate a variation in $\sigma_{8}$.

\begin{table}[htbp]
\centering \protect\protect\protect\protect\caption{\label{tab:t2cosmoerr} $1\sigma$ errors on cosmological parameters
for Type 2 fiducial models.}

\begin{tabular}{l|ccc|c}
\hline 
 & \multicolumn{3}{c|}{FM2-PL} & \multicolumn{1}{c}{FM2-Q}\tabularnewline
Error  & $n=1$  & $n=1.28$  & $n=2$  & \tabularnewline
\hline 
\hline 
$\sigma_{h}$  & 0.035  & 0.035  & 0.029  & 0.031 \tabularnewline
$\sigma_{\Omega_{m}h^{2}}$  & 0.015  & 0.014  & 0.012  & 0.013 \tabularnewline
$\sigma_{\Omega_{b}h^{2}}$  & 0.0033  & 0.032  & 0.0028  & 0.0027 \tabularnewline
$\sigma_{n_{s}}$  & 0.0402  & 0.038  & 0.029  & 0.034 \tabularnewline
$\sigma_{\gamma}$  & 0.036  & 0.035  & 0.036  & 0.052 \tabularnewline
$\sigma_{\sigma_{8}}$  & 0.0049  & 0.0048  & 0.0046  & 0.0073 \tabularnewline
\hline 
\end{tabular}
\end{table}

The errors on the bias parameters $b_{0}$ and $b_{1}$ are listed
in Table~\ref{tab:t2biaserr} for the three power law indices considered.
The trend with $n$ is the same as in the Type 1 case. Errors on $b_{0}$
do not significantly depend on $n$. They do increase with the redshift
(although the relative error $\sigma_{b_{0}}/b_{0}$ does not). Errors
on $b_{1}$ significantly increase with $n$.

\begin{table}[htbp]
\centering \protect\protect\protect\protect\caption{\label{tab:t2biaserr} Errors on bias parameters for Type 2 fiducial
models.}

\begin{tabular}{c|cccccc|ccc}
\hline 
 & \multicolumn{6}{c|}{FM2-PL} & \multicolumn{3}{c}{FM2-Q}\tabularnewline
 & \multicolumn{2}{c}{$n=1$} & \multicolumn{2}{c}{$n=1.28$} & \multicolumn{2}{c|}{$n=2$} & \multicolumn{3}{c}{}\tabularnewline
$z$  & $\sigma_{b_{0}}$  & $\sigma_{b_{1}}$  & $\sigma_{b_{0}}$  & $\sigma_{b_{1}}$  & $\sigma_{b_{0}}$  & $\sigma_{b_{1}}$  & $\sigma_{b_{0}}$  & $\sigma_{Q}$  & $\sigma_{A}$ \tabularnewline
\hline 
\hline 
$0.8$  & 0.014  & 0.12  & 0.012  & 0.19  & 0.011  & 0.91  & 0.029  & 2.6  & 0.46 \tabularnewline
$1.0$  & 0.016  & 0.12  & 0.014  & 0.18  & 0.013  & 0.76  & 0.033  & 1.9  & 0.39 \tabularnewline
$1.2$  & 0.018  & 0.12  & 0.016  & 0.18  & 0.016  & 0.66  & 0.039  & 1.6  & 0.36 \tabularnewline
$1.4$  & 0.021  & 0.13  & 0.019  & 0.19  & 0.019  & 0.69  & 0.045  & 1.6  & 0.34 \tabularnewline
$1.6$  & 0.024  & 0.16  & 0.022  & 0.24  & 0.023  & 0.89  & 0.049  & 1.8  & 0.37 \tabularnewline
$1.8$  & 0.028  & 0.21  & 0.026  & 0.32  & 0.027  & 1.3  & 0.056  & 2.1  & 0.40 \tabularnewline
\hline 
\end{tabular}
\end{table}

\begin{figure}[htb!]
\centering{}\hspace*{-1cm} %
\begin{tabular}{cc}
\includegraphics[width=0.45\textwidth]{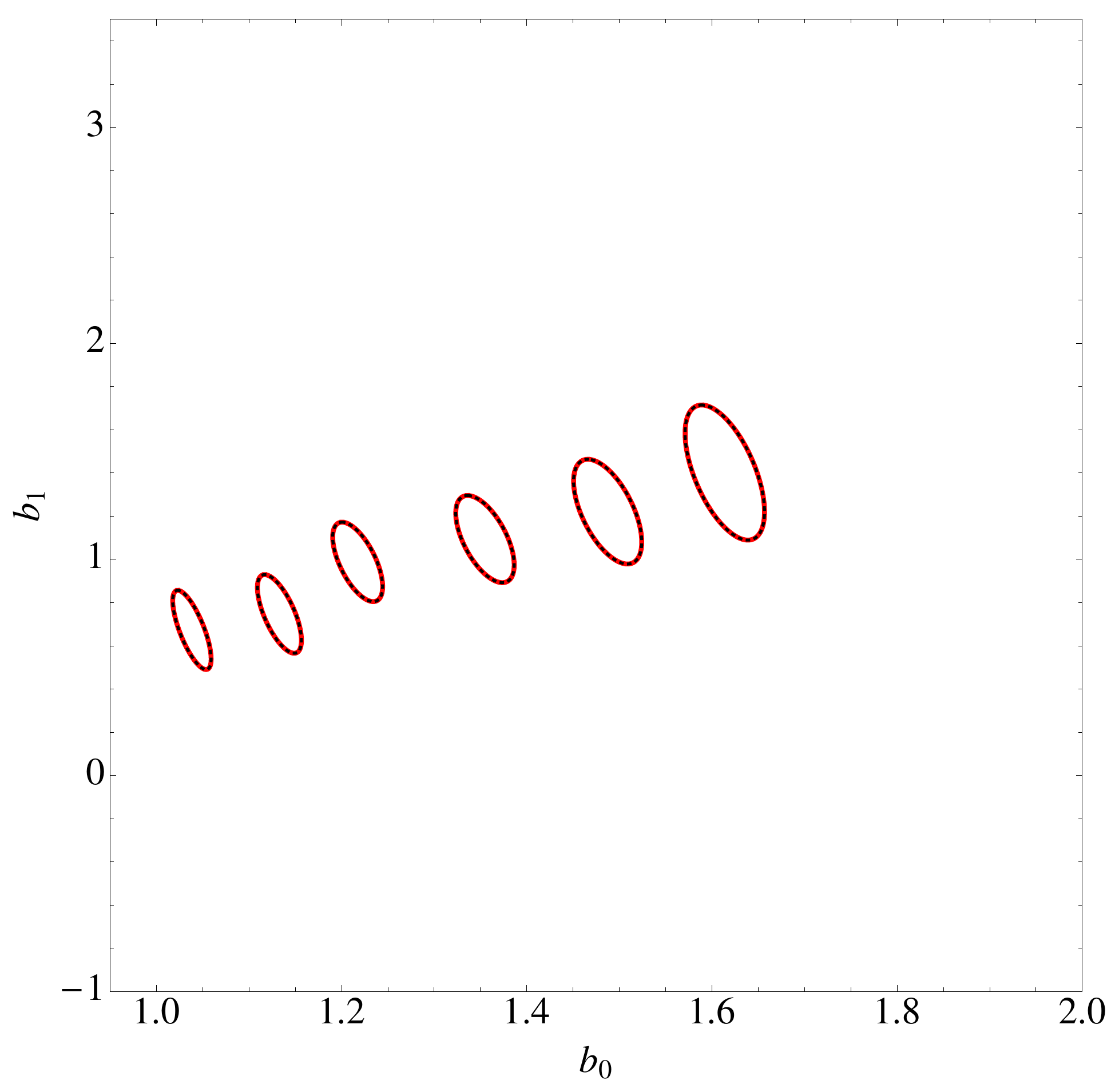}  & \includegraphics[width=0.45\textwidth]{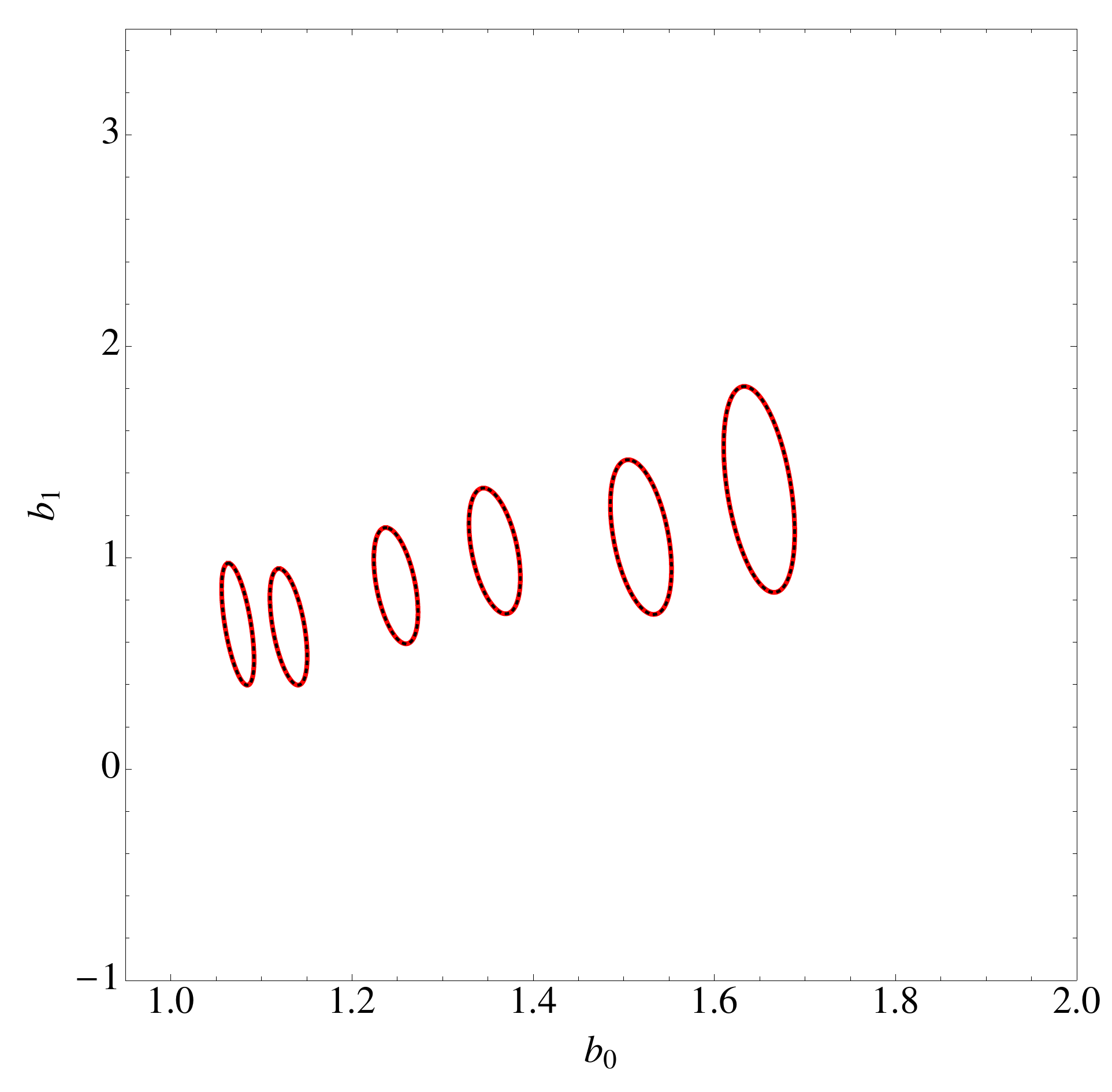}\tabularnewline
\includegraphics[width=0.45\textwidth]{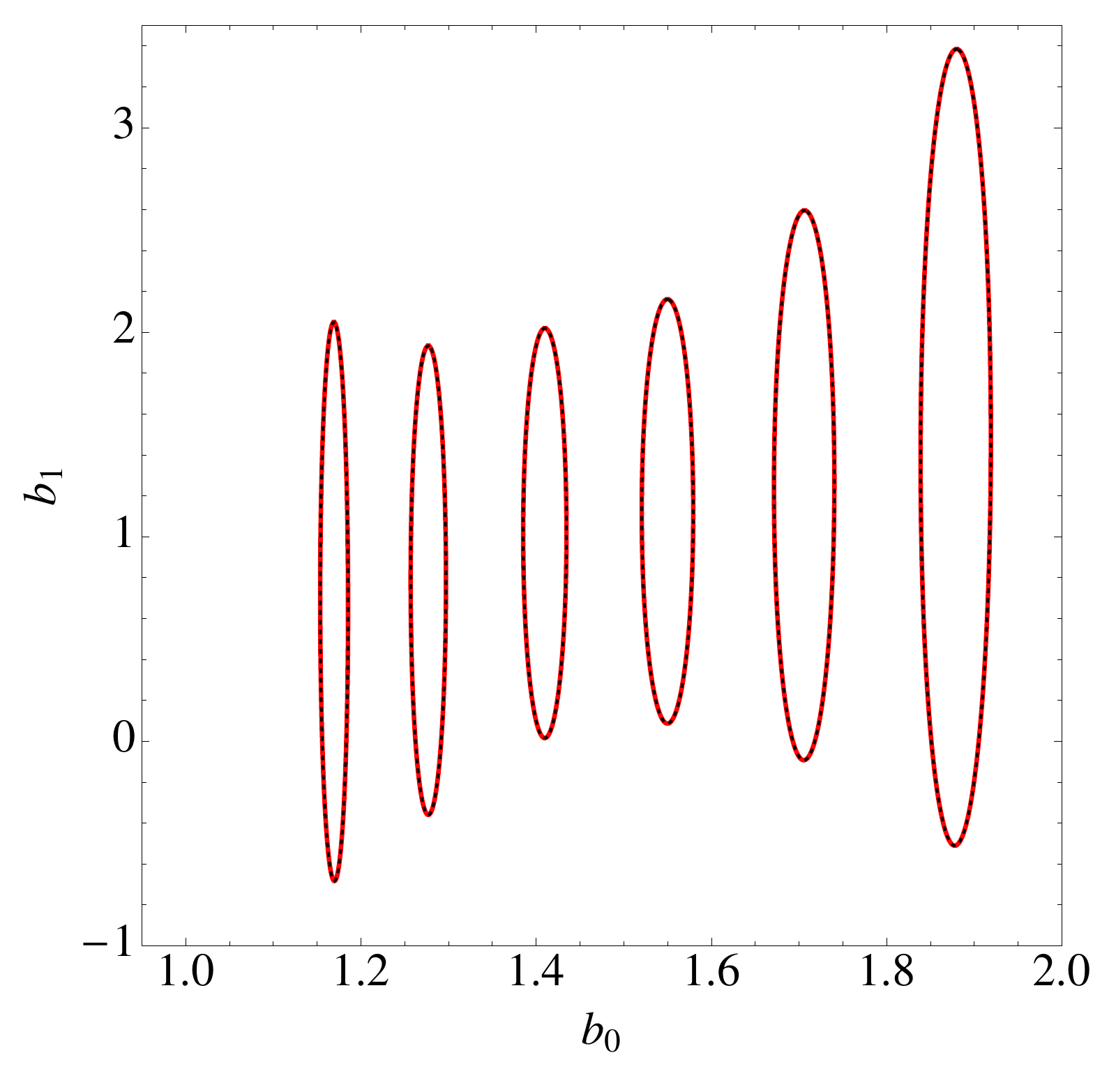}  & \tabularnewline
\end{tabular}\protect\caption{\label{fig:a-powerlaw-euclidb0b1nref1}68 \% probability contours
for the parameters $b_{0}$ and $b_{1}$ of the FM2-PL model for different
values of $n$. They are centred at the $b_{0}$ and $b_{1}$ values
indicated in Tab.~\ref{tab:T2}. Ellipses of increasing redshifts
are centred at increasing values of $b_{0}$. In the left upper panel
we show the case $n=1$. In the right panel we show the $n=1.28$
case. In the low left panel we show the $n=2$ case. Only the redshifts
$z=0.8,1.2,1.6,1.8$ are shown to avoid overcrowding. }
\end{figure}

Fig.~\ref{fig:a-powerlaw-euclidb0b1nref1} illustrates the covariance
between $b_{0}$ and $b_{1}$. The different ellipses are centred
at the best fit values $b_{0}$ and $b_{1}$ listed in Table~\ref{tab:T2}.
As their magnitude steadily increases with the redshift the center
of the corresponding ellipses shifts accordingly.

Errors on $b_{0}$ are small and show a weak dependence on the redshift
and on $n$. Errors on $b_{1}$ are comparatively larger, show a weak
dependence on the redshift and a strong dependence on $n$. One consequence
of this is that the possibility to detect a scale dependent bias depend
on the bias model itself. For $n=1$ the scale dependency, i.e. the
fact that $b_{1}\ne0$, is detected with a significance of more than
$4$$\sigma$ in each redshift bin and to more that $3$$\sigma$
for $n=1.28$, i.e. with the most realistic fiducial model. Instead,
for the extreme case $n=2$ the size of the errors hides the scale
dependency of the bias.

\subsection{Q-model}

For the FM2-Q case the errors on the cosmological parameters, listed
in Table~\ref{tab:t2cosmoerr}, are remarkably similar to those of
the FM2-PL case except for $\sigma_{8}$ and $\gamma$ whose uncertainties
are significantly larger. This is clearly illustrated by the 68\%
error ellipse in Fig.~\ref{fig:s8gamma} (green, dashed curve) that
also confirms the anti-correlation between the two parameters noticed
in the Type 1 analysis. In addition, as in the Type 1 case, the increase
in the error size reflects the fact that the number of free parameters
in the Q-model is larger than in the Power Law model.

\begin{figure}[htb!]
\centering{}\hspace*{-1cm} %
\begin{tabular}{cc}
\includegraphics[width=0.45\textwidth]{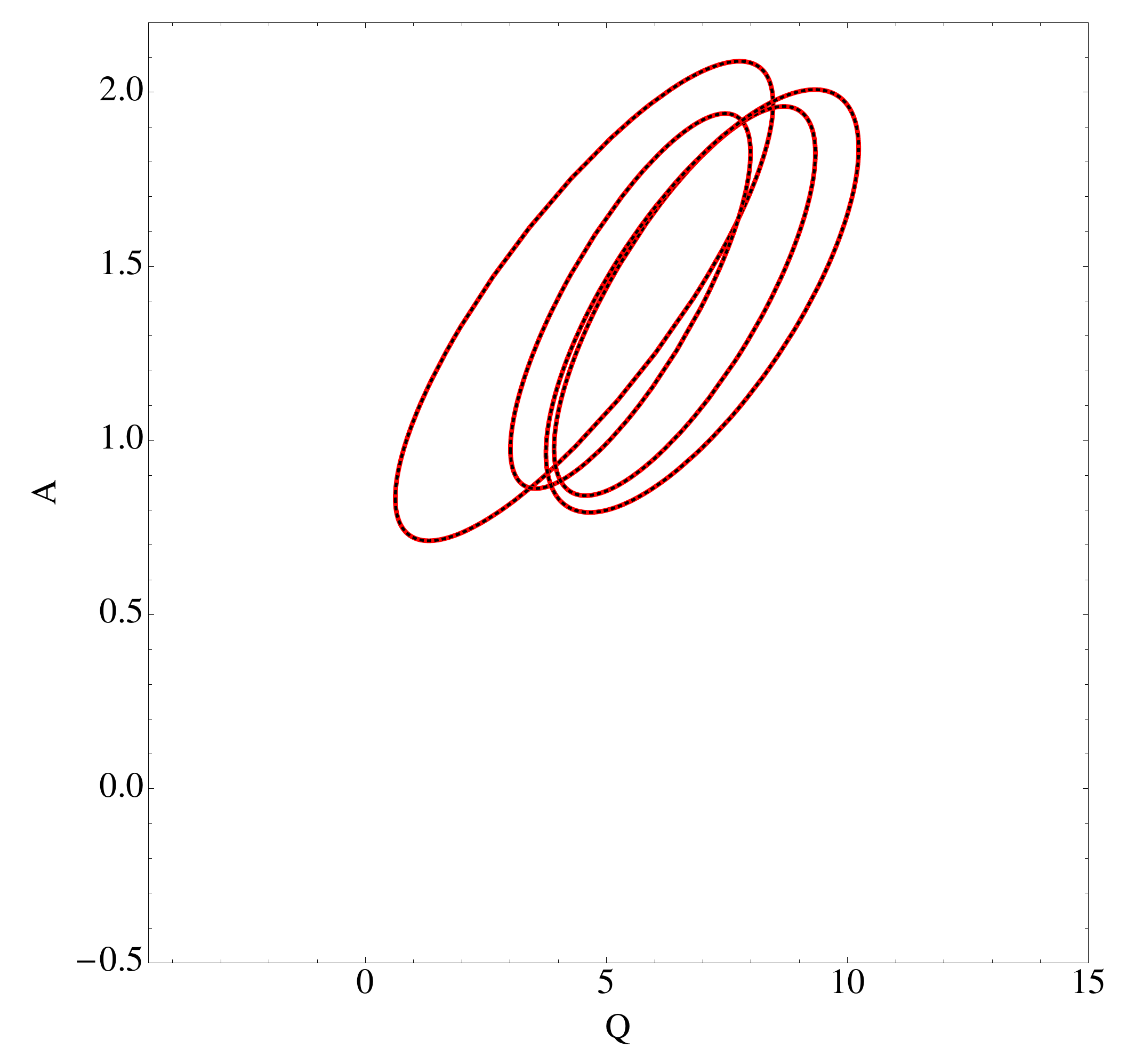}  & \tabularnewline
\end{tabular}\protect\caption{\label{fig:a-euclidcolA0Q}68 \% probability contours for the parameters
$A$ and $Q$ of the FM2-Q model. We plot only the redshifts $z=0.8,1.2,1.6,1.8$,
left to right, to avoid overlaps.}
\end{figure}

The errors on $b_{0}$, $A$ and $Q$ are listed in the last three
columns of Table~\ref{tab:t2biaserr}. Relative errors on $b_{0}$
are similar to those of the FM1-Q case and larger that those of the
FM2-PL case, again reflecting the increase in the number of free parameter
in the bias model. Absolute errors on $A$ and $Q$ are smaller than
in the FM1-Q case and appear to be correlated, as shown in Figure~\ref{fig:a-euclidcolA0Q},
which is expected since the bias in the Q-model depends on their ratio
(\ref{eq:qmodel}). The errors on $A$, $Q$ weakly depend on the
redshift whereas the value of $Q$ significantly increases both at
small and at large redshift (the value of $A$ is set equal to 1.7).
What is most remarkable is the fact that in this realistic fiducial
model the bias parameter $Q$ is significantly ($\sim\,3\sigma$)
different from zero, meaning that the scale dependence in galaxy bias
can be detected even adopting a 3-parameter model. For the parameter
$A$ this significance is smaller and depends on the redshift: it
increases from $\sim\,1\sigma$ at $z=0.8$ to $\sim\,2\sigma$ at
$z=1.8$.

\section{Systematic bias}

Choosing a particular bias fiducial when the galaxy sample is characterized
by another bias model leads to a change in the best fit of the other
parameters. It is interesting then to estimate this ``systematic
bias'' due to the galaxy bias in one of our cases. For simplicity,
we perform this test only for our reference case, the power-law Type
I bias, i.e. FM1-PL $n=1$.

Suppose then we do not estimate the bias from the galaxy clustering
using our method and simply assume the fiducial model FM1-PL, i.e.
we fix the bias at every redshift. This of course produces narrower
errors on the cosmological parameters, but the price to pay is that
the results will depend sensitively on the bias choice. Indeed, if
the chosen bias model is wrong, all the cosmological parameters will
shift by a certain amount. We wish to estimate this shift.

The maximum likelihood estimator $\bar{\theta}_{\alpha}$, given a
likelihood function $L(\theta_{\alpha})$ for the parameters $\theta_{\alpha}$,
is obtained by solving the system of equations 
\begin{equation}
\mathcal{L}_{,\alpha}=0\label{eq:der1}
\end{equation}
where $\mathcal{L}=-\log L$. In the Fisher matrix approximation one
has $\mathcal{L}=\frac{1}{2}D_{i}D_{j}P_{ij}-\frac{1}{2}P$ where
$P=C^{-1}$ is the precision matrix (the inverse of the data correlation
matrix) and $D_{i}=d_{i}-t_{i}$ is the vector of data minus theory
points.

Suppose now $L$ depends also on a parameter $s$ (that we refer to
as a systematic parameter) that has been assigned a particular value,
and we want to estimate how $\bar{\theta}_{\alpha}$ changes when
$s$ is shifted to another value $s+\delta s$, where $\delta s$
is assumed very small. Then we have\emph{ 
\begin{equation}
\mathcal{L}(s+\delta s)=\mathcal{L}(s)+\mathcal{L}_{,s}\delta s
\end{equation}
}and\emph{ }the equations that give the new maximum likelihood estimator
vector $\hat{\theta}=\bar{\theta}+\delta\theta$ become to first order
in $\delta s$ and $\delta\theta$ 
\begin{align}
\mathcal{L}(s+\delta s)_{,\alpha} & =\mathcal{L}_{,\alpha}\mid_{\hat{\theta}}+(\mathcal{L}_{,\alpha s}\mid_{\bar{\theta}})\delta s\\
 & =\mathcal{L}_{,\alpha}\mid_{\bar{\theta}}+(\mathcal{L}_{,\alpha\beta}\mid_{\bar{\theta}})\delta\theta_{\beta}+(\mathcal{L}_{,\alpha s}\mid_{\bar{\theta}})\delta s\\
 & =(\mathcal{L}_{,\alpha\beta}\mid_{\bar{\theta}})\delta\theta_{\beta}+(\mathcal{L}_{,\alpha s}\mid_{\bar{\theta}})\delta s=0
\end{align}
where we employed Eq. (\ref{eq:der1}). Finally we obtain 
\begin{equation}
\delta\theta_{\alpha}=-\mathcal{L}_{,\alpha\beta}^{-1}\mathcal{L}_{,\beta s}\delta s\label{eq:syst}
\end{equation}
(sum over $\beta$). If there are several systematic parameters, then
a sum over the $s$ parameters is understood. Now, once we average
over many data realization, $\langle\mathcal{L}_{,\alpha\beta}\rangle=F_{\alpha\beta}$
is the Fisher matrix, and $\langle\mathcal{L}_{,\beta s}\rangle=F_{\beta s}$
is a sort of systematic Fisher matrix. In practice, this means that
one includes the systematic parameters $s_{i}$ in a general Fisher
matrix that contains also the parameters $\theta_{\alpha}$, and then
selects the $\alpha\beta$ and the $\beta s_{i}$ submatrices and
produces the sum over $\beta$ and $s_{i}$, namely (\cite{1998PhRvL..81.2004K})
\begin{align}
\delta\theta_{\alpha} & =-F_{\alpha\beta}^{-1}F_{\beta s_{i}}\delta s_{i}
\end{align}

We consider separately the case in which one shifts the $b_{0}$ and
$b_{1}$ parameters. We consider separately a systematic change in
$b_{0}$ or $b_{1}$ of $\delta s_{i}=$0.1 in every bin. The results
obviously scale with $\delta s_{i}$. The larger the coefficients
$F_{\alpha\beta}^{-1}F_{\beta s_{i}}$ are, the more sensitive a parameter
is to the determination of the fiducial bias. We see from Table \ref{tab:Shift-of-the}
that the largest effect occurs on $n_{s}$ due to a wrong $b_{0}$,
a shift much larger than the statistical error (compare with Table
\ref{tab:a-referencecosmoerrors}). In general, a wrong choice of
$b_{0}$ of magnitude 0.1 yields a shift on the cosmological parameters
from 2 to 5 times their statistical errors. This is a powerful argument
to insert the bias parameters in the Fisher matrix (and in the data
likelihood when real data will be available) rather than freezing
them to some fiducial value.

We notice that the size of systematic errors is larger for those cosmological
parameters that quantify the amplitude or the overall shape of the
power spectrum and comparatively smaller for those related to sharp
spectral features like $\Omega_{b}$ for BAOs. We conjecture this
will apply also to the characteristic step at the free-streaming scale
induced by massive neutrinos. Finally, a wrong choice of $b_{1}$
(by the same amount of 0.1 in every bin), on the other hand, induces
a shift quite smaller than the statistical error. Allowing for a scale
dependent bias makes in this case little difference.

\begin{table}
\begin{centering}
\begin{tabular}{|c|c|c|}
\hline 
 & $b_{0}$  & $b_{1}$\tabularnewline
\hline 
\hline 
$h$  & -0.12  & 0.0016\tabularnewline
\hline 
$\Omega_{m}h^{2}$  & -0.052  & 0.00088\tabularnewline
\hline 
$\Omega_{b}h^{2}$  & -0.011  & 0.000098\tabularnewline
\hline 
$n_{s}$  & 0.14  & -0.013\tabularnewline
\hline 
$\gamma$  & -0.12  & -0.011\tabularnewline
\hline 
$\sigma_{8}$  & 0.025  & -0.0046\tabularnewline
\hline 
\end{tabular}
\par\end{centering}
\caption{\label{tab:Shift-of-the}Shift of the maximum likelihood parameter
estimates due to a systematic error of 0.1 in the bias parameters
$b_{0}$ and $b_{1}$ in every bin. }
\end{table}

\section{Conclusions}

In this paper we have investigated the impact of a scale-dependent
galaxy bias on the results of the clustering analysis performed in
next generation surveys. We focused on the galaxy power spectrum and,
as a case study, we have considered a spectroscopic redshift survey
with characteristics similar to those expected for the Euclid survey.
In particular we focused on two issues: \textit{i)} the impact of
a scale dependent bias on the estimate of the cosmological parameters
and \textit{ii)} the accuracy with which it will be possible to detect
such scale dependence.

For this purpose we have considered two different models for the scale-dependent
galaxy bias and obtained a Fisher matrix forecast assuming two sets
of fiducial models. Both scale dependent bias models are characterised
by a linear bias parameter $b_{0}$ and additional parameters that
quantify scale dependence. However, in the Power Law model one of
them, the power law index, is fixed, whereas in the Q model both parameters,
$A$ and $Q$, are free too vary. All bias parameters change with
redshift, so they should really be regarded as functions specified
at different redshifts. For as concerns the fiducial models, we have
considered two scenarios. The first one is the simple but unrealistic
case of a population of unbiased mass tracers. The second one is that
of a more realistic galaxy bias relation calibrated on simulated catalogs
of H$\alpha$-line emitting objects mimicking the expected characteristics
of the Euclid survey.

Our main results can be summarised as follows: 
\begin{itemize}
\item Allowing for a scale dependent bias does not increase significantly
the errors on cosmological parameters, except for the growth index
$\gamma$ and the $rms$ density fluctuation $\sigma_{8}$.

More specifically, in our analysis we find that errors on $h_{0}$,
$\Omega_{m0}$, $\Omega_{b0}$, $\Omega_{k0}$, and $n_{s}$ are insensitive
to a scale dependency in galaxy bias, to the specific form of scale-dependent
bias and to the choice of the fiducial model. In fact these errors
are only slightly larger than those expected when one assumes that
bias is scale independent. On the contrary errors on $\gamma$ and
$\sigma_{8}$ are rather sensitive to the bias model and to the fiducial.
In the ideal scenario of a population of unbiased mass tracers the
expected relative error on $\gamma$ is smaller than 5 \%. However,
when a more realistic bias model and survey setup are adopted, the
relative error increases to $\sim6$ \% for a 2-parameter bias model
and to $\sim9$ \% for a 3-parameter model. Clearly, an accurate estimate
of the growth index would benefit from a reliable model for galaxy
bias that can be captured by few free parameters.

In addition, $\gamma$ and $\sigma_{8}$ are correlated. This correlation
is expected since the clustering analysis constrains the amplitude
of the power spectrum, proportional to the product $b(k,z)\sigma_{8}(z)$
and its redshift distortions, which are proportional to $f(z)/b(k,z)$.
The degree of degeneracy, however, depends on the fiducial, being
larger when more realistic bias models are adopted. 
\item The linear bias parameter $b_{0}$ can be determined within a few
\%. The relative error is rather insensitive to the choice of the
fiducial and slightly increases with the redshift. As expected, the
errors increase with the number of free parameters in the model and
therefore is larger in the Q Model than in the Power Law one. Yet,
even in the worst case (the realistic case of a Type 2 Q-Model bias
at $z=1.8$) the relative error is of the order of 2 \%. This is of
the same order as the uncertainty in the estimate of $b_{0}$ obtained
when one assumes no scale dependency, i.e. when one does not consider
derivatives with respect to bias parameters other than $b_{0}$ in
the Fisher matrix analysis (see e.g.\cite{diporto12}). 
\item The accuracy with which one can estimate the bias parameters that
describe the scale dependency depends on the bias model and on the
fiducial. The case in which the scale dependence is stronger, i.e.
the Power Law model with $n=2$, provides the worst results. This
is not surprising since in this case the scale dependence is pushed
at small scales where our Fisher matrix analysis, optimised to probe
linear to mildly nonlinear scales, is less sensitive. However, we
should stress that this $n=2$ case is hardly realistic. In fact this
model provides a poor fit to the bias of H$\alpha$ galaxies in the
simulated Euclid catalog. On the contrary, the Q-model, the Power
Law model with $n=1.28$ and, to a lesser extent, the Power Law model
with $n=1$ match the bias estimated from the mocks.

When we restrict our analysis to the more realistic cases we find
that scale dependency can be clearly detected at all redshifts. The
significance of the detection, quantified by the departures of the
parameters $b_{1}$, $A$ and $Q$ from zero, depends on both the
bias and the fiducial. For the Power Law model with $n=1$ is larger
than $4$$\sigma$ at each redshift, decreasing to $\sim3$$\sigma$
for $n=1.28$. For the Q-model scale dependence is best detected through
the parameter $Q$ ($\sim3$$\sigma$) whereas the significance of
a non zero $A$-value is expected to be smaller and redshift dependent. 
\item An additional benefit in including galaxy bias in the likelihood is
that of reducing the impact of systematic errors. Using a simple argument
based on the ``systematic Fisher matrix\textquotedblright{} approach
(arXiv:1101.1521) we have shown that adopting an incorrect bias model
induces significant systematic errors that, for some cosmological
parameter, can exceed statistical uncertainties. 
\end{itemize}
All in all we conclude that a significant detection of a scale dependent
bias is within reach of next generation redshift surveys and that
contemplating such scale dependence does not significantly decrease
the accuracy in the estimate of most cosmological parameters. We note
that a scale dependent bias can also be induced by primordial non-Gaussianity.
Its signature is expected on much larger scales than those affected
by galaxy bias and there should be ample possibility to disentangle
the two effects. Yet, the need to allow for additional free parameters
to account for non-Gaussian features may have an impact on the precision
of the analysis. We shall explore this issue in a future paper.

\section{Acknowledgements}

The research leading to these results has received funding from the
European Research Council under the European Community' s Seventh
Framework Programme ($FP7/2007-2013$ Grant Agreement no. $279954$),
therefore EM acknowledges the financial support provided by the European
Research Council. EB acknowledges the financial support provided by
NFN-PD51 INDARK, MIUR PRIN 2011 'The dark Universe and the cosmic
evolution of baryons: from current surveys to Euclid' and Agenzia
Spaziale Italiana for financial support from the agreement ASI/INAF/I/023/12/0.
LA acknowledges financial support provided by DFG TRR33 ``The Dark
Universe''. LA acknowledges useful discussions within the Euclid
Theory Working Group. We wish to thank Alex Merson for providing the
100 deg$^{2}$ Euclid light cone mock catalogs.

%\selectlanguage{american}%%\bibliography{observables,bib_master}%\selectlanguage{english}

 \bibliographystyle{plain}
\bibliography{draft_55.bib}

\begin{thebibliography}{10}

\bibitem{theorev}
L.~{Amendola}, S.~{Appleby}, D.~{Bacon}, T.~{Baker}, M.~{Baldi}, N.~{Bartolo},
  A.~{Blanchard}, C.~{Bonvin}, S.~{Borgani}, E.~{Branchini}, C.~{Burrage},
  S.~{Camera}, C.~{Carbone}, L.~{Casarini}, M.~{Cropper}, C.~{de Rham}, C.~{Di
  Porto}, A.~{Ealet}, P.~G. {Ferreira}, F.~{Finelli},
  J.~{Garc{\'{\i}}a-Bellido}, T.~{Giannantonio}, L.~{Guzzo}, A.~{Heavens},
  L.~{Heisenberg}, C.~{Heymans}, H.~{Hoekstra}, L.~{Hollenstein}, R.~{Holmes},
  O.~{Horst}, K.~{Jahnke}, T.~D. {Kitching}, T.~{Koivisto}, M.~{Kunz}, G.~{La
  Vacca}, M.~{March}, E.~{Majerotto}, K.~{Markovic}, D.~{Marsh}, F.~{Marulli},
  R.~{Massey}, Y.~{Mellier}, D.~F. {Mota}, N.~{Nunes}, W.~{Percival},
  V.~{Pettorino}, C.~{Porciani}, C.~{Quercellini}, J.~{Read}, M.~{Rinaldi},
  D.~{Sapone}, R.~{Scaramella}, C.~{Skordis}, F.~{Simpson}, A.~{Taylor},
  S.~{Thomas}, R.~{Trotta}, L.~{Verde}, F.~{Vernizzi}, A.~{Vollmer}, Y.~{Wang},
  J.~{Weller}, and T.~{Zlosnik}.
\newblock {Cosmology and Fundamental Physics with the Euclid Satellite}.
\newblock {\em Living Reviews in Relativity}, 16:6, September 2013.

\bibitem{alhambra13}
P.~{Arnalte-Mur}, V.~J. {Mart{\'{\i}}nez}, P.~{Norberg},
  A.~{Fern{\'a}ndez-Soto}, B.~{Ascaso}, A.~I. {Merson}, J.~A.~L. {Aguerri},
  F.~J. {Castander}, L.~{Hurtado-Gil}, C.~{L{\'o}pez-Sanjuan}, A.~{Molino},
  A.~{Montero-Dorta}, M.~{Stefanon}, E.~{Alfaro}, T.~{Aparicio-Villegas},
  N.~{Ben{\'{\i}}tez}, T.~{Broadhurst}, J.~{Cabrera-Ca{\~n}o}, J.~{Cepa},
  M.~{Cervi{\~n}o}, D.~{Crist{\'o}bal-Hornillos}, A.~{del Olmo}, R.~M.
  {Gonz{\'a}lez-Delgado}, C.~{Husillos}, L.~{Infante}, I.~{M{\'a}rquez},
  J.~{Masegosa}, M.~{Moles}, J.~{Perea}, M.~{Povi{\'c}}, F.~{Prada}, and J.~M.
  {Quintana}.
\newblock {The ALHAMBRA survey: evolution of galaxy clustering since \$z
  $\backslash$sim 1\$}.
\newblock {\em ArXiv e-prints}, November 2013.

\bibitem{bp07}
S.~{Basilakos}, M.~{Plionis}, K.~{Kova{\v c}}, and N.~{Voglis}.
\newblock {Large-scale structure in the HI Parkes All-Sky Survey: filling the
  voids with HI galaxies?}
\newblock {\em \mnras}, 378:301--308, June 2007.

\bibitem{bower06}
R.~G. {Bower}, A.~J. {Benson}, R.~{Malbon}, J.~C. {Helly}, C.~S. {Frenk}, C.~M.
  {Baugh}, S.~{Cole}, and C.~G. {Lacey}.
\newblock {Breaking the hierarchy of galaxy formation}.
\newblock {\em \mnras}, 370:645--655, August 2006.

\bibitem{branchini01}
E.~{Branchini}.
\newblock {Probing the Mass Distribution with IRAS Galaxies}.
\newblock {\em ArXiv Astrophysics e-prints}, October 2001.

\bibitem{coil06}
A.~L. {Coil}, J.~A. {Newman}, M.~C. {Cooper}, M.~{Davis}, S.~M. {Faber}, D.~C.
  {Koo}, and C.~N.~A. {Willmer}.
\newblock {The DEEP2 Galaxy Redshift Survey: Clustering of Galaxies as a
  Function of Luminosity at z = 1}.
\newblock {\em \apj}, 644:671--677, June 2006.

\bibitem{cole05}
S.~{Cole}, W.~J. {Percival}, J.~A. {Peacock}, P.~{Norberg}, C.~M. {Baugh},
  C.~S. {Frenk}, I.~{Baldry}, J.~{Bland-Hawthorn}, T.~{Bridges}, R.~{Cannon},
  M.~{Colless}, and C.~{Collins}.
\newblock {The 2dF Galaxy Redshift Survey: power-spectrum analysis of the final
  data set and cosmological implications}.
\newblock {\em \mnras}, 362:505--534, September 2005.

\bibitem{comparat13}
J.~{Comparat}, E.~{Jullo}, J.-P. {Kneib}, C.~{Schimd}, H.~{Shan}, T.~{Erben},
  O.~{Ilbert}, J.~{Brownstein}, A.~{Ealet}, S.~{Escoffier}, B.~{Moraes},
  N.~{Mostek}, J.~A. {Newman}, M.~E.~S. {Pereira}, F.~{Prada}, D.~J.
  {Schlegel}, D.~P. {Schneider}, and C.~H. {Brandt}.
\newblock {Stochastic bias of colour-selected BAO tracers by joint
  clustering-weak lensing analysis}.
\newblock {\em \mnras}, 433:1146--1160, August 2013.

\bibitem{contreras013}
S.~{Contreras}, C.~{Baugh}, P.~{Norberg}, and N.~{Padilla}.
\newblock {How robust are predictions of galaxy clustering?}
\newblock {\em ArXiv e-prints}, January 2013.

\bibitem{dalal}
N.~{Dalal}, O.~{Dor{\'e}}, D.~{Huterer}, and A.~{Shirokov}.
\newblock {Imprints of primordial non-Gaussianities on large-scale structure:
  Scale-dependent bias and abundance of virialized objects}.
\newblock {\em \prd}, 77(12):123514, June 2008.

\bibitem{diPorto}
C.~{di Porto}, L.~{Amendola}, and E.~{Branchini}.
\newblock {Growth factor and galaxy bias from future redshift surveys: a study
  on parametrizations}.
\newblock {\em \mnras}, 419:985--997, January 2012.

\bibitem{diporto12}
C.~{Di Porto}, L.~{Amendola}, and E.~{Branchini}.
\newblock {Simultaneous constraints on bias, normalization and growth index
  through power spectrum measurements}.
\newblock {\em \mnras}, 423:L97--L101, June 2012.

\bibitem{diporto14}
C.~{Di Porto}, E.~{Branchini}, J.~{Bel}, F.~{Marulli}, M.~{Bolzonella},
  O.~{Cucciati}, S.~{de la Torre}, B.~R. {Granett}, L.~{Guzzo}, C.~{Marinoni},
  L.~{Moscardini}, U.~{Abbas}, C.~{Adami}, S.~{Arnouts}, D.~{Bottini},
  A.~{Cappi}, J.~{Coupon}, I.~{Davidzon}, G.~{De Lucia}, A.~{Fritz},
  P.~{Franzetti}, M.~{Fumana}, B.~{Garilli}, O.~{Ilbert}, A.~{Iovino},
  J.~{Krywult}, V.~{Le Brun}, O.~{Le Fevre}, D.~{Maccagni}, K.~{Malek}, H.~J.
  {McCracken}, L.~{Paioro}, M.~{Polletta}, A.~{Pollo}, M.~{Scodeggio}, L.~A.~M.
  {Tasca}, R.~{Tojeiro}, D.~{Vergani}, A.~{Zanichelli}, A.~{Burden},
  A.~{Marchetti}, D.~{Martizzi}, Y.~{Mellier}, R.~C. {Nichol}, J.~A. {Peacock},
  W.~J. {Percival}, M.~{Viel}, M.~{Wolk}, and G.~{Zamorani}.
\newblock {The VIMOS Public Extragalactic Redshift Survey (VIPERS). Measuring
  nonlinear galaxy bias at z\~{}0.8}.
\newblock {\em ArXiv e-prints}, June 2014.

\bibitem{ewt98}
D.~J. {Eisenstein}, W.~{Hu}, and M.~{Tegmark}.
\newblock {Cosmic Complementarity: H 0 and Omega M from Combining Cosmic
  Microwave Background Experiments and Redshift Surveys}.
\newblock {\em \apjl}, 504:L57, September 1998.

\bibitem{2007ApJ...664..660E}
D.~J. {Eisenstein}, H.-J. {Seo}, and M.~{White}.
\newblock {On the Robustness of the Acoustic Scale in the Low-Redshift
  Clustering of Matter}.
\newblock {\em \apj}, 664:660--674, August 2007.

\bibitem{fkp}
H.~A. {Feldman}, N.~{Kaiser}, and J.~A. {Peacock}.
\newblock {Power-spectrum analysis of three-dimensional redshift surveys}.
\newblock {\em \apj}, 426:23--37, May 1994.

\bibitem{fg93}
J.~N. {Fry} and E.~{Gaztanaga}.
\newblock {Biasing and hierarchical statistics in large-scale structure}.
\newblock {\em Ap. J.}, 413:447--452, August 1993.

\bibitem{gaztanaga05}
E.~{Gazta{\~n}aga}, P.~{Norberg}, C.~M. {Baugh}, and D.~J. {Croton}.
\newblock {Statistical analysis of galaxy surveys - II. The three-point galaxy
  correlation function measured from the 2dFGRS}.
\newblock {\em \mnras}, 364:620--634, December 2005.

\bibitem{giannantonio}
T.~{Giannantonio}, C.~{Porciani}, J.~{Carron}, A.~{Amara}, and A.~{Pillepich}.
\newblock {Constraining primordial non-Gaussianity with future galaxy surveys}.
\newblock {\em \mnras}, 422:2854--2877, June 2012.

\bibitem{grossi}
M.~{Grossi}, L.~{Verde}, C.~{Carbone}, K.~{Dolag}, E.~{Branchini},
  F.~{Iannuzzi}, S.~{Matarrese}, and L.~{Moscardini}.
\newblock {Large-scale non-Gaussian mass function and halo bias: tests on
  N-body simulations}.
\newblock {\em \mnras}, 398:321--332, September 2009.

\bibitem{hoekstra02}
H.~{Hoekstra}, L.~{van Waerbeke}, M.~D. {Gladders}, Y.~{Mellier}, and H.~K.~C.
  {Yee}.
\newblock {Weak Lensing Study of Galaxy Biasing}.
\newblock {\em \apj}, 577:604--614, October 2002.

\bibitem{huff07}
E.~{Huff}, A.~E. {Schulz}, M.~{White}, D.~J. {Schlegel}, and M.~S. {Warren}.
\newblock {Simulations of baryon oscillations}.
\newblock {\em Astroparticle Physics}, 26:351--366, January 2007.

\bibitem{jullo12}
E.~{Jullo}, J.~{Rhodes}, A.~{Kiessling}, J.~E. {Taylor}, R.~{Massey},
  J.~{Berge}, C.~{Schimd}, J.-P. {Kneib}, and N.~{Scoville}.
\newblock {COSMOS: Stochastic Bias from Measurements of Weak Lensing and Galaxy
  Clustering}.
\newblock {\em \apj}, 750:37, May 2012.

\bibitem{kayo04}
I.~{Kayo}, Y.~{Suto}, R.~C. {Nichol}, J.~{Pan}, I.~{Szapudi}, A.~J. {Connolly},
  J.~{Gardner}, B.~{Jain}, G.~{Kulkarni}, T.~{Matsubara}, R.~{Sheth}, A.~S.
  {Szalay}, and J.~{Brinkmann}.
\newblock {Three-Point Correlation Functions of SDSS Galaxies in Redshift
  Space: Morphology, Color, and Luminosity Dependence}.
\newblock {\em \pasj}, 56:415--423, June 2004.

\bibitem{1998PhRvL..81.2004K}
L.~{Knox}, R.~{Scoccimarro}, and S.~{Dodelson}.
\newblock {Impact of Inhomogeneous Reionization on Cosmic Microwave Background
  Anisotropy}.
\newblock {\em Physical Review Letters}, 81:2004--2007, September 1998.

\bibitem{kovac011}
K.~{Kova{\v c}}, C.~{Porciani}, S.~J. {Lilly}, C.~{Marinoni}, L.~{Guzzo},
  O.~{Cucciati}, G.~{Zamorani}, A.~{Iovino}, P.~{Oesch}, M.~{Bolzonella},
  Y.~{Peng}, and B.~{Meneux}.
\newblock {The Nonlinear Biasing of the zCOSMOS Galaxies up to z \~{} 1 from
  the 10k Sample}.
\newblock {\em \apj}, 731:102, April 2011.

\bibitem{euclidRB}
R.~{Laureijs}, J.~{Amiaux}, S.~{Arduini}, J.~. {Augu{\`e}res}, J.~{Brinchmann},
  R.~{Cole}, M.~{Cropper}, C.~{Dabin}, L.~{Duvet}, A.~{Ealet}, and et~al.
\newblock {Euclid Definition Study Report}.
\newblock {\em ArXiv e-prints}, October 2011.

\bibitem{lewis00}
A.~{Lewis}, A.~{Challinor}, and A.~{Lasenby}.
\newblock {Efficient Computation of Cosmic Microwave Background Anisotropies in
  Closed Friedmann-Robertson-Walker Models}.
\newblock {\em \apj}, 538:473--476, August 2000.

\bibitem{marinoni05}
C.~{Marinoni}, O.~{Le F{\`e}vre}, B.~{Meneux}, A.~{Iovino}, A.~{Pollo},
  O.~{Ilbert}, G.~{Zamorani}, L.~{Guzzo}, and A.~{Mazure}.
\newblock {The VIMOS VLT Deep Survey. Evolution of the non-linear galaxy bias
  up to z = 1.5}.
\newblock {\em \aap}, 442:801--825, November 2005.

\bibitem{marulli13}
F.~{Marulli}, M.~{Bolzonella}, E.~{Branchini}, I.~{Davidzon}, S.~{de la Torre},
  B.~R. {Granett}, L.~{Guzzo}, and A.~{Iovino}.
\newblock {The VIMOS Public Extragalactic Redshift Survey (VIPERS). Luminosity
  and stellar mass dependence of galaxy clustering at 0.5<z<1.1}.
\newblock {\em ArXiv e-prints}, March 2013.

\bibitem{nishimichi07}
T.~{Nishimichi}, I.~{Kayo}, C.~{Hikage}, K.~{Yahata}, A.~{Taruya}, Y.~P.
  {Jing}, R.~K. {Sheth}, and Y.~{Suto}.
\newblock {Bispectrum and Nonlinear Biasing of Galaxies: Perturbation Analysis,
  Numerical Simulation, and SDSS Galaxy Clustering}.
\newblock {\em \pasj}, 59:93--106, February 2007.

\bibitem{norberg02}
P.~{Norberg}, C.~M. {Baugh}, E.~{Hawkins}, S.~{Maddox}, D.~{Madgwick},
  O.~{Lahav}, S.~{Cole}, C.~S. {Frenk}, I.~{Baldry}, J.~{Bland-Hawthorn},
  T.~{Bridges}, R.~{Cannon}, M.~{Colless}, C.~{Collins}, W.~{Couch},
  G.~{Dalton}, R.~{De Propris}, S.~P. {Driver}, G.~{Efstathiou}, R.~S. {Ellis},
  K.~{Glazebrook}, C.~{Jackson}, I.~{Lewis}, S.~{Lumsden}, J.~A. {Peacock},
  B.~A. {Peterson}, W.~{Sutherland}, and K.~{Taylor}.
\newblock {The 2dF Galaxy Redshift Survey: the dependence of galaxy clustering
  on luminosity and spectral type}.
\newblock {\em \mnras}, 332:827--838, June 2002.

\bibitem{norberg01}
P.~{Norberg}, C.~M. {Baugh}, E.~{Hawkins}, S.~{Maddox}, J.~A. {Peacock},
  S.~{Cole}, C.~S. {Frenk}, and J.~{Bland-Hawthorn}.
\newblock {The 2dF Galaxy Redshift Survey: luminosity dependence of galaxy
  clustering}.
\newblock {\em \mnras}, 328:64--70, November 2001.

\bibitem{nuza012}
S.~E. {Nuza}, A.~G. {Sanchez}, F.~{Prada}, A.~{Klypin}, D.~J. {Schlegel},
  S.~{Gottloeber}, A.~D. {Montero-Dorta}, and M.~{Manera}.
\newblock {The clustering of galaxies at z\~{}0.5 in the SDSS-III Data Release
  9 BOSS-CMASS sample: a test for the LCDM cosmology}.
\newblock {\em ArXiv e-prints}, February 2012.

\bibitem{schlegel11}
D.~{Schlegel}, F.~{Abdalla}, T.~{Abraham}, C.~{Ahn}, C.~{Allende Prieto},
  J.~{Annis}, E.~{Aubourg}, and M.~{Azzaro}.
\newblock {The BigBOSS Experiment}.
\newblock {\em ArXiv e-prints}, June 2011.

\bibitem{sw06}
A.~E. {Schulz} and M.~{White}.
\newblock {Scale-dependent bias and the halo model}.
\newblock {\em Astroparticle Physics}, 25:172--177, March 2006.

\bibitem{seljak01}
U.~{Seljak}.
\newblock {Redshift-space bias and {$\beta$} from the halo model}.
\newblock {\em \mnras}, 325:1359--1364, August 2001.

\bibitem{seo03}
H.-J. {Seo} and D.~J. {Eisenstein}.
\newblock {Probing Dark Energy with Baryonic Acoustic Oscillations from Future
  Large Galaxy Redshift Surveys}.
\newblock {\em \apj}, 598:720--740, December 2003.

\bibitem{seo05}
H.-J. {Seo} and D.~J. {Eisenstein}.
\newblock {Baryonic Acoustic Oscillations in Simulated Galaxy Redshift
  Surveys}.
\newblock {\em \apj}, 633:575--588, November 2005.

\bibitem{2007ApJ...665...14S}
H.-J. {Seo} and D.~J. {Eisenstein}.
\newblock {Improved Forecasts for the Baryon Acoustic Oscillations and
  Cosmological Distance Scale}.
\newblock {\em \apj}, 665:14--24, August 2007.

\bibitem{simon07}
P.~{Simon}, M.~{Hetterscheidt}, M.~{Schirmer}, T.~{Erben}, P.~{Schneider},
  C.~{Wolf}, and K.~{Meisenheimer}.
\newblock {GaBoDS: The Garching-Bonn Deep Survey. VI. Probing galaxy bias using
  weak gravitational lensing}.
\newblock {\em \aap}, 461:861--879, January 2007.

\bibitem{skibba13}
R.~A. {Skibba}, M.~S.~M. {Smith}, A.~L. {Coil}, J.~{Moustakas}, J.~{Aird},
  M.~R. {Blanton}, A.~D. {Bray}, R.~J. {Cool}, D.~J. {Eisenstein}, A.~J.
  {Mendez}, K.~C. {Wong}, and G.~{Zhu}.
\newblock {PRIMUS: Galaxy Clustering as a Function of Luminosity and Color at
  0.2<z<1}.
\newblock {\em ArXiv e-prints}, October 2013.

\bibitem{smith07}
R.~E. {Smith}, R.~{Scoccimarro}, and R.~K. {Sheth}.
\newblock {Scale dependence of halo and galaxy bias: Effects in real space}.
\newblock {\em \prd}, 75(6):063512, March 2007.

\bibitem{springel05}
V.~{Springel}, S.~D.~M. {White}, A.~{Jenkins}, C.~S. {Frenk}, N.~{Yoshida},
  L.~{Gao}, J.~{Navarro}, R.~{Thacker}, D.~{Croton}, J.~{Helly}, J.~A.
  {Peacock}, S.~{Cole}, P.~{Thomas}, H.~{Couchman}, A.~{Evrard}, J.~{Colberg},
  and F.~{Pearce}.
\newblock {Simulations of the formation, evolution and clustering of galaxies
  and quasars}.
\newblock {\em \nat}, 435:629--636, June 2005.

\bibitem{swanson08}
M.~E.~C. {Swanson}, M.~{Tegmark}, M.~{Blanton}, and I.~{Zehavi}.
\newblock {SDSS galaxy clustering: luminosity and colour dependence and
  stochasticity}.
\newblock {\em \mnras}, 385:1635--1655, April 2008.

\bibitem{teg97}
M.~{Tegmark}.
\newblock {Measuring Cosmological Parameters with Galaxy Surveys}.
\newblock {\em Physical Review Letters}, 79:3806--3809, November 1997.

\bibitem{tegmark99}
M.~{Tegmark} and B.~C. {Bromley}.
\newblock {Observational Evidence for Stochastic Biasing}.
\newblock {\em \apjl}, 518:L69--L72, June 1999.

\bibitem{verde02}
L.~{Verde}, A.~F. {Heavens}, W.~J. {Percival}, S.~{Matarrese}, C.~M. {Baugh},
  J.~{Bland-Hawthorn}, and T.~{Bridges}.
\newblock {The 2dF Galaxy Redshift Survey: the bias of galaxies and the density
  of the Universe}.
\newblock {\em \mnras}, 335:432--440, September 2002.

\bibitem{zehavi05}
I.~{Zehavi}, Z.~{Zheng}, D.~H. {Weinberg}, J.~A. {Frieman}, A.~A. {Berlind},
  M.~R. {Blanton}, R.~{Scoccimarro}, R.~K. {Sheth}, M.~A. {Strauss}, I.~{Kayo},
  Y.~{Suto}, M.~{Fukugita}, O.~{Nakamura}, N.~A. {Bahcall}, J.~{Brinkmann},
  J.~E. {Gunn}, G.~S. {Hennessy}, {\v Z}.~{Ivezi{\'c}}, G.~R. {Knapp},
  J.~{Loveday}, A.~{Meiksin}, D.~J. {Schlegel}, D.~P. {Schneider},
  I.~{Szapudi}, M.~{Tegmark}, M.~S. {Vogeley}, D.~G. {York}, and {SDSS
  Collaboration}.
\newblock {The Luminosity and Color Dependence of the Galaxy Correlation
  Function}.
\newblock {\em \apj}, 630:1--27, September 2005.

\end{thebibliography}
 %draft.bib~}

\end{document}